\documentclass[12pt, peerreview, draftcls, onecolumn, letterpaper]{IEEEtran}
\usepackage{amsmath}
\usepackage{amsfonts}
\usepackage{amssymb}
\usepackage{epsfig}
\usepackage{units}
\usepackage{nicefrac}
\usepackage[noadjust]{cite}
\usepackage{subfigure}
\usepackage{bm}
\usepackage{verbatim}
\usepackage{pstricks, pst-plot, pst-grad, pstricks-add, pst-node, pstricks-add}
\usepackage[printonlyused]{acronym}	    % \ac, \acf (full), \acs (short), \acl (without acronym)
\usepackage{colortbl}

\newcommand{\Bnull}{\mathbf{0}}

\newcommand{\Ba}{\mathbf{a}}

\newcommand{\Bh}{\mathbf{h}}

\newcommand{\BH}{\mathbf{H}}
\newcommand{\BI}{\mathbf{I}}

\newcommand{\BP}{\mathbf{P}}

\newcommand{\Br}{\mathbf{r}}

\newcommand{\BE}{\mathbf{E}}

\newcommand{\Bp}{\mathbf{p}}

\newcommand{\Bpmax}{\Bp^{\textnormal{max}}}

\newcommand{\BPmax}{\BP^{\textnormal{max}}}

\newcommand{\Bn}{\mathbf{n}}

\newcommand{\Bv}{\mathbf{v}}

\newcommand{\Bs}{\mathbf{s}}

\newcommand{\disd}{d_{\textnormal{ISD}}}

\newcommand{\By}{\mathbf{y}}

\newcommand{\BHest}{\hat{\mathbf{H}}}

\newcommand{\BHeff}[1]{\mathbf{H}_{#1}^{\textnormal{e}}}
\newcommand{\hatBHeff}[1]{\hat{\mathbf{H}}_{#1}^{\textnormal{e}}}

\newcommand{\BHeffH}[1]{\left(\mathbf{H}_{#1}^{\textnormal{e}}\right)^H}
\newcommand{\hatBHeffH}[1]{\left(\hat{\mathbf{H}}_{#1}^{\textnormal{e}}\right)^H}

\newcommand{\BEeff}{\mathbf{E}^{\textnormal{e}}}

\newcommand{\barBEeff}{\bar{\mathbf{E}}^{\textnormal{e}}}

\newcommand{\barBEeffm}{\bar{\mathbf{E}}^{\textnormal{e}}_{m}}

\newcommand{\Bheff}[1]{\mathbf{h}_{#1}^{\textnormal{e}}}
\newcommand{\BheffH}[1]{\left(\mathbf{h}_{#1}^{\textnormal{e}}\right)^H}

\newcommand{\eeffij}{e^{\textnormal{e}}_{i,j}}
\newcommand{\heffij}{h^{\textnormal{e}}_{i,j}}
\newcommand{\hatheffij}{\hat{h}^{\textnormal{e}}_{i,j}}

\newcommand{\bareeffij}{\bar{e}^{\textnormal{e}}_{i,j}}

\newcommand{\inv}[1]{\left(#1\right)^{-1}}
\newcommand{\logdet}[1]{\textnormal{log}_2\left|#1\right|}
\newcommand{\arr}[2]{\left[\begin{array}{#1}#2\end{array}\right]}

\newcommand{\Piim}{\mathbf{\Phi}^{\textnormal{ii}}_{m}}

\newcommand{\Pnn}{\mathbf{\Phi}^{\textnormal{nn}}}

\newcommand{\Pqq}{\mathbf{\Phi}^{\textnormal{qq}}}

\newcommand{\Phh}{\mathbf{\Phi}^{\textnormal{vv}}}

\newcommand{\Pyy}{\mathbf{\Phi}^{\textnormal{yy}}}

\newcommand{\barPyy}{\bar{\mathbf{\Phi}}^{\textnormal{yy}}}

\newcommand{\LOG}{\textnormal{log}}

\newcommand{\MAX}{\textnormal{max}}

\newcommand{\DIAG}{\textnormal{diag}}

\newcommand{\dg}{\Delta}
\newcommand{\SPA}{\textnormal{ }}

\newcommand{\Nbs}{N_{\textnormal{bs}}}

\newcommand{\Nsym}{N_{\textnormal{sym}}}

\newcommand{\NBS}{N_{\textnormal{BS}}}

\newcommand{\Npilots}{N_{\textnormal{p}}}
\newcommand{\Np}{N_{\textnormal{p}}}

\newcommand{\ppilots}{p_{\textnormal{pilots}}}

\newcommand{\SK}{\mathcal{K}}

\newcommand{\SM}{\mathcal{M}}
\newcommand{\SF}{\mathcal{F}}

\newcommand{\SFall}{\mathcal{\SF}_{\textnormal{all}}}

\newcommand{\SX}{\mathcal{X}}

\newcommand{\SZ}{\mathcal{Z}}

\newcommand{\SP}{\mathcal{P}}

\newcommand{\RRyz}{\mathcal{R}^{\textnormal{yz}}}
\newcommand{\RRulzero}{\mathcal{R}_{0}}

\newcommand{\RRulinf}{\mathcal{R}_{\infty}}

\newcommand{\RRuldis}{\mathcal{R}^{\textnormal{dis}}}

\newcommand{\RRulcif}{\mathcal{R}^{\textnormal{cif}}}

\newcommand{\RRuldasc}{\mathcal{R}^{\textnormal{dasc}}}
\newcommand{\RRuldasd}{\mathcal{R}^{\textnormal{dasd}}}

\newcommand{\Ful}{F}

\newcommand{\Xul}{X}

\definecolor{areagray}{rgb}{0.95, 0.95, 0.95}
\definecolor{areadarkgray}{rgb}{0.8, 0.8, 0.8}

\definecolor{arealightgreen}{rgb}{0.9, 1, 0.9}
\definecolor{areagreen}{rgb}{0.7, 0.9, 0.7}
\definecolor{areadarkgreen}{rgb}{0.5, 0.8, 0.5}

\definecolor{arealightblue}{rgb}{0.9, 0.9, 1}
\definecolor{areablue}{rgb}{0.7, 0.7, 0.9}
\definecolor{areadarkblue}{rgb}{0.5, 0.5, 0.8}

\definecolor{arealightred}{rgb}{1, 0.9, 0.9}
\definecolor{areared}{rgb}{0.9, 0.7, 0.7}
\definecolor{areadarkred}{rgb}{0.8, 0.5, 0.5}

\definecolor{arealightpurple}{rgb}{1, 0.9, 1}
\definecolor{areapurple}{rgb}{0.9, 0.7, 0.9}
\definecolor{areadarkpurple}{rgb}{0.8, 0.5, 0.8}

\definecolor{arealightbrown}{rgb}{1, 0.95, 0.9}
\definecolor{areabrown}{rgb}{0.9, 0.8, 0.7}
\definecolor{areadarkbrown}{rgb}{0.8, 0.75, 0.5}

\definecolor{mypurple}{rgb}{0.6, 0, 0.6}
\definecolor{mybrown}{rgb}{1, 0.6, 0}
\definecolor{grey}{rgb}{0.5, 0.5, 0.5}

\newcommand{\ShowFigureMain}[1]{~}

\newtheorem{theorem}{Theorem}

\newrgbcolor{darkgreen}{0.2 0.6 0.2}
\newrgbcolor{darkred}{0.6 0.2 0.2}
\newrgbcolor{verylightgray}{0.85 0.85 0.85}
\newrgbcolor{phase1}{0.2 0.6 0.2}
\newrgbcolor{phase2}{0.6 0.2 0.2}

\usepackage[english]{babel}

\acrodef{CDMA}[CDMA]{code-division multiple access}
\acrodef{CF}[CF]{compress-and-forward}
\acrodef{DF}[DF]{decode-and-forward}
\acrodef{OFDM}[OFDM]{orthogonal frequency division multiplex}
\acrodef{SINR}[SINR]{signal-to-interference-and-noise ratio}
\acrodef{SNR}[SNR]{signal-to-noise ratio}
\acrodef{CSIR}[CSIR]{channel state information at the receiver}
\acrodef{CSI}[CSI]{channel state information}
\acrodef{UE}[UE]{user equipment}
\acrodef{BS}[BS]{base station}
\acrodef{LTE}[LTE]{long term evolution}
\acrodef{MAC}[MAC]{multiple access channel}
\acrodef{BC}[BC]{broadcast channel}
\acrodef{DIS}[DIS]{distributed interference subtraction}
\acrodef{CIF}[CIF]{compressed interference forwarding}
\acrodef{DASC}[DAS-C]{distributed antenna system - centralized}
\acrodef{DASD}[DAS-D]{distributed antenna system - decentralized}
\acrodef{SIC}[SIC]{successive interference cancellation}
\acrodef{RRH}[RRH]{remote radio head}
\acrodef{CoMP}[CoMP]{coordinated multi-point}
\acrodef{IC}[IC]{interference channel}
\acrodef{ARPU}[ARPU]{average revenue per user}
\acrodef{LTE}[LTE]{long term evolution}
\acrodef{CEO}[CEO]{central estimation officer}
\acrodef{SISO}[SISO]{single-input single-output}
\acrodef{MIMO}[MIMO]{multiple-input multiple-output}
\acrodef{MMSE}[MMSE]{minimum mean square error}
\acrodef{TDM}[TDM]{time division multiplex}
\acrodef{FDM}[FDM]{frequency division multiplex}
\acrodef{SPC}[SPC]{superposition coding}
\acrodef{MRC}[MRC]{maximum ratio combining}
\acrodef{IRC}[IRC]{interference rejection combining}
\acrodef{DASN}[DAS-N]{network DAS}

\title{Uplink CoMP under a Constrained Backhaul and Imperfect Channel Knowledge}
\author{
  Patrick~Marsch,~\IEEEmembership{Member,~IEEE,}~and~Gerhard~Fettweis,~\IEEEmembership{Fellow,~IEEE}%
  \thanks{Manuscript submitted \today. Part of this work has been published in the proceedings of the IEEE International Conference on Communications (ICC) 2009~\cite{Marsch_ICC09}.}}% ZIH should be mentioned
\pubid{~}
\begin{document}
\maketitle

  \begin{abstract}
  Coordinated Multi-Point (CoMP) is known to be a key technology for next generation mobile communications systems, as it allows to overcome the burden of inter-cell interference. Especially in the uplink, it is likely that interference exploitation schemes will be used in the near future, as they can be used with legacy terminals and require no or little changes in standardization. Major drawbacks, however, are the extent of additional backhaul infrastructure needed, and the sensitivity to imperfect channel knowledge. This paper jointly addresses both issues in a new framework incorporating a multitude of proposed theoretical uplink CoMP concepts, which are then put into perspective with practical CoMP algorithms. This comprehensive analysis provides new insight into the potential usage of uplink CoMP in next generation wireless communications systems.
  \end{abstract}

\begin{keywords}
CoMP, network MIMO, constrained backhaul, imperfect CSI, joint detection, interference cancellation, multiple access channel, interference channel
\end{keywords}

\newpage

\section{Introduction}
\label{s:INTRODUCTION}

\subsection{Motivation}
\label{s:INTRODUCTION_MOTIVATION}

Mobile network operators are experiencing an exponentially growing demand for mobile data rates at a stagnating \ac{ARPU}, driving the need for larger {\em spectral efficiency}. It is known, however, that especially urban cellular systems are mainly limited through inter-cell interference~\cite{GuptaKumar_IEEETRANS00}. To overcome this limitation, \ac{CoMP} was proposed in~\cite{BaierWeber_ISSSTA00, ShamaiZaidel_VTC01}, and has been selected as a key technology for \ac{LTE}-Advanced~\cite{Parkvall_JC09}. In the uplink, for instance, multi-cell joint signal processing enables the exploitation of interference~\cite{SklavosWeber_Kluwer02, ShamaiZaidel_JWCC04}, rather than treating it as noise, promising vast gains in spectral efficiency and fairness~\cite{Andrews_IEEEWCOM05,MarschKhattak_ITW06}. Beside key challenges, such as synchronization in time and frequency, a major concern of uplink \ac{CoMP} is its demand for additional backhaul~\cite{Marsch_ICC07}, and its sensitivity to imperfect \ac{CSI}~\cite{Marsch_ICC09}.

This paper performs an analysis of various uplink \ac{CoMP} concepts under a constrained {\em out-of-band} backhaul and imperfect \ac{CSI}. The joint observation of these two major issues from a theoretical and practical perspective sheds a new light on the value of particular \ac{CoMP} schemes in next generation wireless communication systems. 

\subsection{Related Work}
\label{s:INTRODUCTION_RELATEDWORK}
  
Considering the aspect of a \textbf{constrained backhaul}, an uplink \ac{CoMP} scenario is related to the CEO-problem~\cite{ViswanathanBerger_IEEETRANS97}, where a number of agents make noisy, but correlated observations on the same random source, and use capacity-constrained links to a \ac{CEO}, who aims at reconstructing the source with minimum distortion. For a Gaussian source and noise, and a quadratic distortion measure, the rate-distortion trade-off was found in~\cite{Oohama_IEEETRANS98}, respectively.

In~\cite{SanderovichShamai_ISIT05}, transmission from a two-antenna \ac{UE} to two \acp{BS} linked to a central processing unit was considered as a particular \ac{CEO} problem setup. The work was based on {\em distributed Wyner-Ziv compression}~\cite{Gastpar_IEEETRANS04}, though its optimality could not yet be proved. The work was extended in~\cite{CosoSimoens,delCosoSimoens_ISIT08} to the case of multiple \ac{UE}s, pointing out that compression can trade-off one \ac{UE}'s rate versus the other, and to an arbitrary number of \ac{BS}s with symmetric inter-cell interference in a circular Wyner model in~\cite{SanderovichShamai_ISIT07, ShamaiPoor_JWCC07, ShamaiZaidel_PIMRC08, SanderovichShamai_IEEETRANS09}.

While previous citations considered the exchange of quantized receive signals and {\em centralized} decoding, it can be beneficial under strongly constrained backhaul to use {\em decentralized} decoding where the \ac{BS}s exchange decoded data bits~\cite{Marsch_WCNC08, SimeoneShamai_ASILOMAR08, SimeoneShamai_IEEETRANS09}, or quantizations of transmit sequences~\cite{SimeoneShamai_IEEETRANS09, GriegerMarsch_GLOBECOM09} for (partial) interference subtraction. The benefit of adapting between different cooperation strategies depending on the channel realization has been pointed out in~\cite{Marsch_WCNC08, Marsch_WPMC08}.

Concrete \ac{CoMP} {\em algorithms} have been proposed using centralized~\cite{KhattakRave_IST06, MayerHagenauer_ICC06} or decentralized decoding~\cite{BavarianCavers_GLOBECOM07, AktasHanly_IEEETRANS08}, where the latter schemes involve an iterative exchange of likelihood information on transmitted bits. Considering the overall rate/backhaul trade-off, however, iterative schemes are only marginally superior to single-shot cooperation~\cite{MayerHagenauer_ICC06, GriegerFettweis_PIMRC09}. In general, each \ac{BS} may only exchange information connected to its own \ac{UE}s~\cite{KhattakFettweis_VTC07}, or also that on interfering \ac{UE}s~\cite{KhattakFettweis_VTC08}, and the rate/backhaul trade-off strongly depends on the quantization approach~\cite{KhattakFettweis_EURASIP08}.

A different perspective on backhaul-constrained uplink \ac{CoMP} is to see the setting as an \ac{IC} with partial receiver-side cooperation~\cite{WangTse_Imprint09}, distinguishing between gains in {\em degrees of freedom}~\cite{EtkinTse_CORR07} and in {\em power}. However, the cited work also considers scenarios of {\em strong interference} (acc. to~\cite{EtkinTse_CORR07}) that are unlikely to occur in the context of cellular systems, as the assignment of \ac{UE}s to \ac{BS}s would simply be swapped on a reasonable time basis, and excludes the important option of centralized multi-user decoding.

Considering the aspect of \textbf{imperfect \ac{CSI}}, first information theoretical steps concerning the impact on \ac{SISO} links were taken in~\cite{Medard_IEEETRANS00}, and extended to point-to-point \ac{MIMO} links in~\cite{YooGoldsmith_IEEETRANS06}. The impact on uplink \ac{CoMP} was studied from a signal processing perspective in~\cite{MeurerWeber_ITGSCC04, WeberSklavos_IEEETRANS06} and in information theory in~\cite{Marsch_ICC09}. 

%Note that some authors consider \ac{BS} cooperation to take place over the same wireless resource as the communication between \ac{UE}s and \ac{BS}s~\cite{HostMadsen_IEEETRANS06, PrabhakaranViswanath_CORR09}. In this work, however, we constrain ourselves to {\em out-of band} signaling between \ac{BS}s, assuming a dedicated backhaul infrastructure.

\subsection{Main Contribution of this Work}
\label{s:INTRODUCTION_MAINCONTRIBUTION}

This work yields new conclusions on uplink \ac{CoMP} in practical systems, providing
\begin{itemize}
\item a framework incorporating a multitude of information theoretic concepts provided by various authors, putting these in perspective with a variety of proposed signal processing schemes. 
\item numerical results considering both information theoretic bounds as well as practical constraints, and hence yielding an insight into the value of sophisticated signal processing. 
\item reasonably complex models reflecting interference scenarios likely to occur in practical cellular systems. While these models do not enable closed-form analysis, they yield more relevant conclusions than overly simplified models as, e.g., in~\cite{SanderovichShamai_ISIT07, ShamaiPoor_JWCC07, ShamaiZaidel_PIMRC08}.
\end{itemize}

\subsection{Terminology}
\label{s:INTRODUCTION_TERMINOLOGY}
  
In this work, the terms {\em CoMP} and {\em \ac{BS} cooperation} refer to schemes where \ac{BS}s exchange received signals or information connected to the data bits of certain \ac{UE}s in order to improve data rates. Schemes that only make use of {\em coordination} between \ac{BS}s, for example joint scheduling or \ac{IRC}, are considered {\em non-cooperative}. The term {\em backhaul infrastructure} refers to the overall connectivity of \ac{BS}s and the network, while any {\em backhaul} quantity always refers to the backhaul capacity required by a cooperative scheme {\em in addition} to that of a non-cooperative system.
 
\subsection{Outline}
  
In Section~\ref{s:BASICS}, the transmission model and basic \ac{BS} cooperation schemes are introduced, inner bounds on capacity regions under imperfect \ac{CSI} for infinite, no, or partial \ac{BS} cooperation are derived, and performance regions are introduced. In Section~\ref{s:ANALYSIS}, the overall \ac{CoMP} gain is quantified for different scenarios, and the introduced \ac{BS} cooperation schemes are evaluated w.r.t. the achievable rate/backhaul trade-off. The value of \ac{BS} cooperation in conjunction with source coding or superposition coding is discussed, before Monte Carlo simulations using a slightly larger setup emphasize the gain of adaptation between different \ac{BS} cooperation strategies. In Section~\ref{s:PRACTICAL}, parallels are drawn between the analyzed theoretical concepts and proposed practical algorithms, and the value of iterative \ac{BS} cooperation and other practical aspects are discussed. The work is concluded in Section~\ref{s:CONCLUSIONS}.

\section{System Model and Basics}
\label{s:BASICS}

%In this section, some basic terminology is explained, after which the considered transmission model is described. It is shown how the impact of imperfect \ac{CSIR} is modeled. Then, inner bounds on the capacity regions of our setup with infinite or no BS cooperation are derived. Finally, four different base station cooperation schemes are introduced, for which lower bounds on the sum-rate are derived. The section is completed with an introduction of the concept of performance regions.  
    
\subsection{Transmission Model}
\label{s:BASICS_TXMODEL} 

\begin{figure}
\begin{center}
\begingroup
\unitlength=1mm
\begin{picture}(140, 90)(0, 0)
  %\put(0, 0){\framebox(88, 55)}

  \psset{xunit=1mm, yunit=1mm, linewidth=0.2mm}

  % Place the terminals
  \rput(20, 5){\cnodeput[fillstyle=solid, fillcolor=lightgray](0, 0){UE1}{\scalebox{0.7}{UE $1$}}}
	\rput(60, 5){\cnodeput[fillstyle=solid, fillcolor=lightgray](0, 0){UE2}{\scalebox{0.7}{UE $2$}}}
	\rput(90, 5){$\cdots$}
	\rput(120, 5){\cnodeput[fillstyle=solid, fillcolor=lightgray](0, 0){UEK}{\scalebox{0.7}{UE $K$}}}
	
  % Place the multiplication signs
  \rput(20, 20){\rnode{mult1}{$\otimes$}}
  \rput(60, 20){\rnode{mult2}{$\otimes$}}
	\rput(120, 20){\rnode{multK}{$\otimes$}}

  % place the channel vectors
  \rput(12, 20){\rnode{h1}{$\Bh_1$}}
  \rput(52, 20){\rnode{h2}{$\Bh_2$}}
  \rput(112, 20){\rnode{hK}{$\Bh_K$}}
  
  % Place the addition signs
  \rput(10, 50){\rnode{plus11}{$\oplus$}}
  \rput(30, 50){\rnode{plus12}{$\oplus$}}
  \rput(50, 50){\rnode{plus21}{$\oplus$}}
  \rput(70, 50){\rnode{plus22}{$\oplus$}}
  \rput(110, 50){\rnode{plusM1}{$\oplus$}}
  \rput(130, 50){\rnode{plusM2}{$\oplus$}}
  \rput(20, 57){$\cdots$}
	\rput(60, 57){$\cdots$}
	\rput(120, 57){$\cdots$}
	
  % place the noise terms
  \rput(2, 50){\rnode{n11}{$n_{1,1}$}}
  \rput(20, 50){\rnode{n12}{$n_{1,\Nbs}$}}
  \rput(42, 50){\rnode{n21}{$n_{2,1}$}}
  \rput(60, 50){\rnode{n22}{$n_{2,\Nbs}$}}
  \rput(101, 50){\rnode{nM1}{$n_{M,1}$}}
  \rput(119, 50){\rnode{nM2}{$n_{M,\Nbs}$}}
  
  % Place the base stations
  \rput(10, 65){\rnode{BS11}{\psframebox[linestyle=none]{\white B}}}
  \rput(30, 65){\rnode{BS12}{\psframebox[linestyle=none]{\white B}}}
  \rput(20, 65){\rnode{BS1}{\psframebox[fillstyle=solid, fillcolor=lightgray]{Base station $1$}}}
  \rput(50, 65){\rnode{BS21}{\psframebox[linestyle=none]{\white B}}}
  \rput(70, 65){\rnode{BS22}{\psframebox[linestyle=none]{\white B}}}
  \rput(60, 65){\rnode{BS2}{\psframebox[fillstyle=solid, fillcolor=lightgray]{Base station $2$}}}
  \rput(110, 65){\rnode{BSM1}{\psframebox[linestyle=none]{\white B}}}
  \rput(130, 65){\rnode{BSM2}{\psframebox[linestyle=none]{\white B}}}
  \rput(120, 65){\rnode{BSM}{\psframebox[fillstyle=solid, fillcolor=lightgray]{Base station $M$}}}
  \rput(90, 65){$\cdots$}
  
  % Place the network
  \rput(70, 80){\rnode{NW}{\psframebox[linestyle=dashed, linecolor=gray]{\gray Network}}}
  
  % draw connections
  \ncline[linestyle=dotted]{->}{mult1}{plus11}
  \ncline[linestyle=dotted]{->}{mult2}{plus11}
  \ncline[linestyle=dotted]{->}{multK}{plus11}
  \ncline[linestyle=dotted]{->}{mult1}{plus12}
  \ncline[linestyle=dotted]{->}{mult2}{plus12}
  \ncline[linestyle=dotted]{->}{multK}{plus12}
  \ncline[linestyle=dotted]{->}{mult1}{plus21}
  \ncline[linestyle=dotted]{->}{mult2}{plus21}
  \ncline[linestyle=dotted]{->}{multK}{plus21}
  \ncline[linestyle=dotted]{->}{mult1}{plus22}
  \ncline[linestyle=dotted]{->}{mult2}{plus22}
  \ncline[linestyle=dotted]{->}{multK}{plus22}
  \ncline[linestyle=dotted]{->}{mult1}{plusM1}
  \ncline[linestyle=dotted]{->}{mult2}{plusM1}
  \ncline[linestyle=dotted]{->}{multK}{plusM1}
  \ncline[linestyle=dotted]{->}{mult1}{plusM2}
  \ncline[linestyle=dotted]{->}{mult2}{plusM2}
  \ncline[linestyle=dotted]{->}{multK}{plusM2}

  % signals from UEs into channel
  \ncline[linestyle=solid]{->}{UE1}{mult1}\mput*{$s_1$}
  \ncline[linestyle=solid]{->}{UE2}{mult2}\mput*{$s_2$}
  \ncline[linestyle=solid]{->}{UEK}{multK}\mput*{$s_K$}

  % multiplied channel coefficients
  \ncline[linestyle=solid]{->}{h1}{mult1}
  \ncline[linestyle=solid]{->}{h2}{mult2}
  \ncline[linestyle=solid]{->}{hK}{multK}

  % added noise terms
  \ncline[linestyle=solid]{->}{n11}{plus11}
  \ncline[linestyle=solid]{->}{n12}{plus12}
  \ncline[linestyle=solid]{->}{n21}{plus21}
  \ncline[linestyle=solid]{->}{n22}{plus22}
  \ncline[linestyle=solid]{->}{nM1}{plusM1}
  \ncline[linestyle=solid]{->}{nM2}{plusM2}
  
  % signals going into BSs
  \ncline[linestyle=solid]{->}{plus11}{BS11}\mput*{$y_{1,1}$}
  \ncline[linestyle=solid]{->}{plus12}{BS12}\mput*{$y_{1,\Nbs}$}
  \ncline[linestyle=solid]{->}{plus21}{BS21}\mput*{$y_{2,1}$}
  \ncline[linestyle=solid]{->}{plus22}{BS22}\mput*{$y_{2,\Nbs}$}
  \ncline[linestyle=solid]{->}{plusM1}{BSM1}\mput*{$y_{M,1}$}
  \ncline[linestyle=solid]{->}{plusM2}{BSM2}\mput*{$y_{M,\Nbs}$}

  % backhaul connections
	\ncarc[linestyle=solid, linecolor=lightgray, linewidth=2pt]{-}{BS1}{BS2}
	\ncarc[linestyle=solid, linecolor=lightgray, linewidth=2pt, arcangleA=16, arcangleB=16]{-}{BS1}{BSM}
	\ncarc[linestyle=dotted, linecolor=lightgray, linewidth=2pt]{-}{BS1}{NW}
	\ncarc[linestyle=solid, linecolor=lightgray, linewidth=2pt]{-}{BS2}{BSM}
	\ncarc[linestyle=dotted, linecolor=lightgray, linewidth=2pt]{-}{BS2}{NW}
	\ncarc[linestyle=dotted, linecolor=lightgray, linewidth=2pt]{-}{NW}{BSM}

  % place channel and backhaul comment
  \rput(70, 73){\psframebox[fillstyle=solid, fillcolor=white, linestyle=none]{Backhaul infrastructure}}
  \rput(70, 35){\psframebox[fillstyle=solid, fillcolor=white, linestyle=none]{Channel}}
      
  \end{picture}
\endgroup
\end{center}
\caption{Uplink transmission considered in this work.}
\label{f:BASICS_SETUP}
\end{figure}
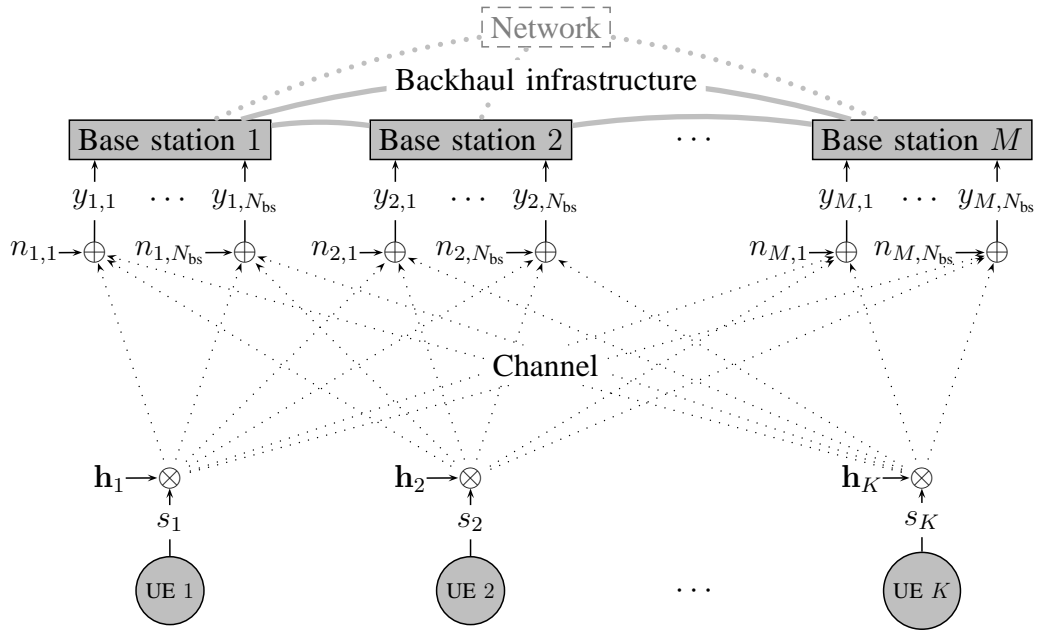

We consider an uplink transmission from $K$ \ac{UE}s to $M$ \ac{BS}s, as shown in Fig.~\ref{f:BASICS_SETUP}, and denote the sets of \ac{UE}s and \ac{BS}s as $\SK=\{1..K\}$ and $\SM=\{1..M\}$, respectively. We assume that each \ac{UE} has $N_{\textnormal{ue}}=1$ transmit antenna, as this is the configuration in the recently completed standard \ac{LTE} Release 8~\cite{McCoy_2007}. The BSs can be equipped with any number $\Nbs$ of receive antennas each. We assume that transmission takes place over a frequency-flat channel, where all entities are perfectly synchronized in time and frequency. Each \ac{UE} $k \in \SK$ has a set $\SF_k$ of discrete messages which it maps onto a set $\SX_k$ of Gaussian unit power transmit sequences of length $\Nsym$ symbols, using an encoding function $e(\cdot)$. We denote all messages of all \ac{UE}s as $\SFall = \SF_1 \cup \SF_2 \cup \cdots \cup \SF_K$, and all transmit sequences as $\SX_{\textnormal{all}} = \SX_1 \cup \SX_2 \cup \cdots \cup \SX_K$. The overall transmission from all \ac{UE}s to the \ac{BS}s in one single channel access $1 \leq t \leq \Nsym$ is given as
\begin{equation}
\label{e:BASICS_TRANSMISSION}
\By^{[t]} = \BH \Bs^{[t]} + \Bn^{[t]},
\end{equation}

\noindent where \scalebox{0.9}{$\By^{[t]} = [y_{1,1}^{[t]}..y_{1, \Nbs}^{[t]}, y_{2,1}^{[t]}..y_{2, \Nbs}^{[t]}, \cdots, y_{M,1}^{[t]}..y_{M,\Nbs}^{[t]}]^T \in \mathbb{C}^{[\NBS \times 1]}$} are the signals received at the BSs, and the channel matrix is given as \scalebox{0.9}{$\BH = [\Bh_1 \Bh_2 \cdots \Bh_K] \in \mathbb{C}^{[\NBS \times K]}$}, where each column $\Bh_k$ is connected to UE $k$. The channel is assumed to be block-fading, where each element is taken from an independent, zero-mean Gaussian distribution $h_{i,j} \sim \mathcal{N}_{\mathbb{C}}\left(0, E\left\{ |h_{i,j}|^2 \right\} \right)$. $\Bs^{[t]} \in \mathbb{C}^{[K \times 1]}$ are the symbols transmitted from the UEs, which are given as
\begin{equation}
\label{e:BASICS_TRANSMITTED_SIGNALS}
\forall \SPA k \in \SK : \SPA s_k^{[t]} = \sum\limits_{\forall \SPA F \in \SF_k} \sqrt{\rho_{F}} \left[e(F)\right]^{[t]},
\end{equation}

\noindent where $\rho_{F} \in \mathbb{R}_0^+$ is the transmit power assigned to message $F$. We use $\SP = \{\rho_{F} : \SPA F \in \SFall \}$ to capture the overall power allocation. According to~\eqref{e:BASICS_TRANSMITTED_SIGNALS}, each UE $k$ transmits a weighted superposition of sequences in set $\SX_k$. \scalebox{0.9}{$\Bn^{[t]} = [n_{1,1}^{[t]}..n_{1, \Nbs}^{[t]}, n_{2,1}^{[t]}..n_{2, \Nbs}^{[t]}, \cdots, n_{M,1}^{[t]}..n_{M,\Nbs}^{[t]}]^T \in \mathbb{C}^{[\NBS \times 1]}$} is additive Gaussian noise at the receiver side with covariance $E_t\{\Bn^{[t]}(\Bn^{[t]})^H\} = \sigma^2 \BI$. We state the covariance of the transmitted signals as $E_t\{\Bs^{[t]}(\Bs^{[t]})^H\} = \BP = \DIAG(\Bp)$ with $\Bp \in \mathbb{R}_0^{+[K \times 1]}$, where each element $p_k$ corresponds to the overall transmit power (over all transmitted sequences) of UE $k$, and assume that the transmit powers are subject to the power constraint $\BPmax - \BP \succeq 0$. Hence, each UE has an individual power constraint defined by the entries of the diagonal matrix $\BPmax = \DIAG(\Bpmax)$ with \scalebox{0.9}{$\Bpmax \in \mathbb{R}_0^{+[K \times 1]}$}. The transmit covariance connected to only a subset of messages $\SF \subseteq \SF_{\textnormal{all}}$ is denoted as $\BP(\SF)$, where the diagonal elements are given as
\begin{equation}
\label{e:BASICS_SUBSETPOWERS}
\left[\BP(\SF)\right]_{k,k} = \sum\limits_{\forall \SPA F \in \left(\SF_k \cap \SF\right)} \rho_{F}.
\end{equation}

We also use $\forall \SPA k \in \SK : \SPA S_k = \{s_k^{[1]}..s_k^{[\Nsym]}\}$ as the superposition of all sequences transmitted by \ac{UE} $k$, $\forall \SPA m \in \SM : \SPA Y_m$ as the sequence of all symbols received at all antennas of \ac{BS} $m$, and $\forall \SPA m \in \SM, 1 \leq a \leq \Nbs: \SPA N_{m,a}$ as the noise sequence received by BS $m$ at antenna $a$. As indicated in Fig.~\ref{f:BASICS_SETUP}, the \ac{BS}s are assumed to be connected through a mesh of error-free out-of-band backhaul links, where we denote as $\beta \in \mathbb{R}^+_0$ the sum backhaul capacity required in addition to that of a non-cooperative system. Note that in our setup, it is sufficient if a \ac{UE} can be decoded by {\em any} involved \ac{BS}, which then forwards the decoded bits to the network, circumventing cases of strong interference~\cite{EtkinTse_CORR07}. The symbol index $t$ is omitted in the sequel for brevity.

\subsection{Modeling of Imperfect Channel Knowledge}
\label{s:BASICS_IMPCSI}

To incorporate the impact of imperfect (receiver-side) \ac{CSI} into our model, we assume that all \ac{BS}s have the same knowledge of the compound channel estimate
\begin{equation}
\label{e:BASICS_CHNEST}
\BHest = \BH + \BE,
\end{equation}

\noindent with $\BHest \in \mathbb{C}^{[\NBS \times K]}$, where the error term $\BE \in \mathbb{C}^{[\NBS \times K]}$ is a random variable of covariance
\begin{equation}
\label{e:BASICS_KRAMER}
E\left\{vec(\BE)vec(\BE)^H\right\} = \frac {\sigma_{\textnormal{pilots}}^2} {\Npilots \cdot p_{\textnormal{pilots}}} \cdot \BI = \sigma_E^2 \cdot \BI.
\end{equation}
The estimated channel $\BHest$ and error $\BE$ are assumed to have multiple independent realizations per block of $\Nsym$ symbols (due to multiple pilots per block). Equation~\eqref{e:BASICS_KRAMER} is based on the Kramer-Rao lower bound~\cite{Kay_BOOK93}, yielding the absolute estimation error variance, if optimal channel estimation has been performed based on $\Npilots$ pilots of power $\ppilots$, subject to Gaussian noise with variance $\sigma_{\textnormal{pilots}}^2$. Note that $\sigma_{\textnormal{pilots}}^2$ can differ from $\sigma^2$ if multi-cell (quasi-)orthogonal pilot sequences are employed. In the sequel, we assume unit-power pilots ($p_{\textnormal{pilots}} = 1$), and choose $\Np = 2$, which has been motivated through the observation of a concrete channel estimation scheme in a frequency-selective OFDMA system for a channel of average coherence time and bandwidth in~\cite{MarschRostFettweis_WSA10}. Let us now state the following theorem:
\begin{theorem}[Modified transmission equation under imperfect CSI]
\label{t:BASICSCHNEST_EQTRANS}
An inner bound for the capacity region (considering {\em average rates} over many estimation errors) of the transmission in~\eqref{e:BASICS_TRANSMISSION} under imperfect \ac{CSI} can be found by observing the capacity region connected to the transmission
\begin{equation}
\label{e:BASICS_EFFECTIVE_TRANSMISSION}
\By = \BHeff{} \Bs + \Bv + \Bn,
\end{equation} 
\noindent which involves a power-reduced effective channel $\BHeff{} \in \mathbb{C}^{[\NBS \times K]}$ with elements
\begin{equation}
\label{e:BASICS_EFFECTIVE_CHANNEL}
\forall \SPA i,j : \SPA \heffij = \frac {h_{i,j}} {\sqrt{1 + \sigma_E^2 \left/ E\left\{\left|h_{i,j}\right|^2\right\}\right.}},
\end{equation}

\noindent and is subject to an additional Gaussian noise term $\Bv \in \mathbb{C}^{[\NBS \times 1]}$ with diagonal covariance 
\begin{equation}
\label{e:BASICS_EFFECTIVE_NOISE}
E\left\{\Bv \Bv^H \right\} = \Phh = \dg \left( \barBEeff \BP \left( \SFall \right) \left(\barBEeff\right)^H \right), \SPA \textnormal{ where } \SPA \forall \SPA i,j : \bareeffij = \sqrt{\frac {E\left\{\left|h_{i,j}\right|^2\right\} \cdot \sigma_E^2}{E\left\{\left|h_{i,j}\right|^2\right\} + \sigma_E^2}},
\end{equation}

\noindent and $\dg(\cdot)$ sets all off-diagonal values of the operand to zero.
\end{theorem}

\begin{proof}
Briefly, the theorem is based on the fact that~\eqref{e:BASICS_EFFECTIVE_TRANSMISSION} overestimates the detrimental impact of imperfect \ac{CSI} by assuming $\Bv$ to be a Gaussian random variable with a different realization in each channel use. The proof is stated in the Appendix.
\end{proof}

Note that our model of the channel estimate in~\eqref{e:BASICS_CHNEST} deviates from that in, e.g.,~\cite{YooGoldsmith_IEEETRANS06}, where the authors start with the assumption of an {\em unbiased} \ac{MMSE} channel estimate which is uncorrelated from its estimation error. Both models, however, lead to Theorem~\ref{t:BASICSCHNEST_EQTRANS}, while the model considered in this work has the advantage that $\sigma_E^2$ is given as an {\em absolute} channel estimation noise term, where the different impact on weak or strong links becomes evident in~\eqref{e:BASICS_EFFECTIVE_NOISE}. In general, the model implies that $\sigma_E^2$, as well as the average gain of all links $E\{|h_{i,j}|^2\}$, are known to the receiver side.
 
%\begin{figure*}[b]
%\hrulefill
%\begin{equation}
%\label{e:COND_COVARIANCE}
%\Pyy^{\mathcal{M}|m^*} = \BHeff^{\mathcal{M}} \left( \BI + \BP \BHeff^{m^*} \left( \Pnn^{m^*} + \Phh^{m^*} \right)^{-1} (\BHeff^{m^*})^H \right) \BP \left(\BHeff^{\mathcal{M}}\right)^H + \Phh^{\mathcal{M}} + \Pnn^{\mathcal{M}}
%\end{equation}
%\end{figure*}

\subsection{Capacity Region Under Infinite BS Cooperation}
\label{s:BASICS_INF}

If an infinite backhaul infrastructure enables full cooperation between all \ac{BS}s, we are observing a \ac{MAC}. In this context, there is no benefit of superimposed messages~\cite{CoverThomas_BOOK06}, hence we can constrain the used messages to
\begin{equation}
\label{s:BASICS_INF_MESSAGESETS}
\forall \SPA k \in \SK : \SPA \SF_k := \left\{F_k\right\}, \SPA \SF_{\textnormal{all}} := \left\{F_1, F_2, \cdots, F_K \right\} \SPA \textnormal{and} \SPA \SP := \left\{\rho_{F_1}, \rho_{F_2}, \cdots, \rho_{F_K}\right\}
\end{equation}

\noindent and state the following theorem:
\begin{theorem}[Capacity region under infinite BS cooperation]
\label{t:BASICS_INF_CAPREGION}
An inner bound for the capacity region of the uplink transmission in~\eqref{e:BASICS_TRANSMISSION} under infinite \ac{BS} cooperation is given as
\begin{equation}
\label{e:BASICS_INF_CAPREGION_GENERAL}
\RRulinf = \bigcup_{\SP \SPA : \SPA \BPmax - \BP\left(\SF_{\textnormal{all}}\right) \succeq 0} \RRulinf(\SP) 
\end{equation}

\noindent where $\bigcup$ denotes a convex hull operation, and all rate tuples $\Br \in \RRulinf(\SP)$ fulfill $\forall \SPA k \in \SK : 0 \leq r_k \leq \nu_{F_k}$ and $\forall \SPA \SF \subseteq \SF_{\textnormal{all}}$:  
\begin{equation}
\label{e:BASICS_INF_CAPREGION_GIVENP}
\sum\limits_{F \in \SF} \nu_{F} \leq \LOG_2 \left| \BI + \left( \sigma^2 \BI + \Phh \right)^{-1} \BHeff{} \BP\left(\SF \right) \BHeffH{} \right|,
\end{equation}

\noindent where $\nu_{F}$ is the rate connected to message $F$.

\end{theorem}
\begin{proof}
The proof is a straightforward application of~\cite{Telatar_IEEETRANS99} to~\eqref{e:BASICS_EFFECTIVE_TRANSMISSION} and given in~\cite{Marsch_DISS10}.
\end{proof}

Equation~\eqref{e:BASICS_INF_CAPREGION_GIVENP} states that the sum rate of any subset of \ac{UE}s is limited by the sum capacity of the channel, assuming that all other \ac{UE}s have already been decoded and their signals subtracted from the system. Note that under imperfect \ac{CSI}, a certain extent of noise covariance $\Phh$ remains, having a detrimental impact on any cooperation strategy that will be explored later. If the sum rate is to be maximized and all links have equal average power, \eqref{e:BASICS_INF_CAPREGION_GIVENP} simplifies to the expression derived for {\em point-to-point MIMO transmission} under imperfect \ac{CSI} in~\cite{YooGoldsmith_IEEETRANS06}. 

\subsection{Capacity Region without BS Cooperation}
\label{s:BASICS_ZERO}

Without \ac{BS} cooperation, our scenario is similar to a Gaussian \ac{IC}, where the capacity is known only for certain interference cases. The tightest known inner bound~\cite{HanKobayashi_IEEETRANS81} is based on \ac{SPC}, where {\em common messages} are decoded individually by multiple receivers. Our setup differs in the way that we can swap the assignment of \ac{UE}s to \ac{BS}s, or let \ac{BS}s decode multiple \ac{UE}s, such that scenarios of {\em strong interference}~\cite{EtkinTse_CORR07} are avoided. This reduces the range of scenarios for which common message concepts are known to be beneficial, and the increased background noise level due to imperfect \ac{CSI} renders these even less attractive. As \ac{SPC} in general suffers from \ac{SNR} gaps inherent in practical coding schemes, and requires \ac{UE} modifications in conjunction with more complex signaling, we again constrain ourselves to one message per \ac{UE} as in~\eqref{s:BASICS_INF_MESSAGESETS}. Term $\Ba \in \{1..M\}^{[K \times 1]}$ captures \ac{BS}-\ac{UE} assignment, i.e. denotes the \ac{BS} where each \ac{UE} is decoded at, and we introduce
\begin{equation}
\label{e:BASICS_ZERO_SETSOFDECODEDMESSAGES}
\forall \SPA m \in \SM : \SPA \SF^{[m]}\left(\Ba\right) = \left\{ \Ful_k \in \SF_{\textnormal{all}} : a_k = m \right\} \SPA \textnormal{and} \SPA \bar{\SF}^{[m]}\left(\Ba\right) = \left\{ \Ful_k \in \SF_{\textnormal{all}} : \SPA a_k \neq m \right\}
\end{equation}

\noindent as the sets of messages {\em decoded} or {\em not decoded} by \ac{BS} $m$, respectively. We now state:
\begin{theorem}[Inner bound on the capacity region without BS cooperation]
\label{t:BASICS_ZERO_CAPREGION}
An inner bound of the capacity region of the transmission in~\eqref{e:BASICS_TRANSMISSION} without \ac{BS} cooperation is given as
\begin{equation}
\label{e:BASICS_ZERO_CAPREGION_GENERAL}
\RRulzero = {\bigcup}_{\Ba, \SP \SPA : \SPA \BP\left(\SF_{\textnormal{all}}\right) \SPA \preceq \SPA \BPmax} \RRulzero\left(\Ba, \SP\right),
\end{equation}

\noindent where all rate tuples $\Br \in \RRulzero(\Ba, \SP)$ fulfill $\forall \SPA k \in \SK : \SPA 0 \leq r_k \leq \nu_{\Ful_k}$ and $\forall \SPA m \in \SM$ :
\begin{multline}
\label{e:BASICS_ZERO_CAPREGION_GIVENP}
\forall \SPA \SF' \subseteq \SF^{[m]}\left(\Ba\right) : \sum\limits_{F \in \SF'} \nu_F \leq \LOG_2 \left| \BI + \left( \Piim \right)^{-1} \BHeff{m} \BP\left(\SF'\right) \BHeffH{m} \right| \\
\textnormal{with} \SPA \Piim = \sigma^2 \BI + \underbrace{\Phh}_{\textnormal{Impact of imp. CSI}} + \underbrace{\BHeff{m} \BP\left(\bar{\SF}^{[m]}\left(\Ba\right)\right) \BHeffH{m}}_{\textnormal{Interference}},
\end{multline}

\noindent where $\BHeff{m}$ and $\barBEeffm$ denote effective channel and part of $\barBEeff$ from~\eqref{e:BASICS_EFFECTIVE_NOISE}, resp., connected to \ac{BS} $m$. 

\end{theorem}
\begin{proof}
The theorem is a straightforward extension of the work in~\cite{HanKobayashi_IEEETRANS81} to the transmission in~\eqref{e:BASICS_EFFECTIVE_TRANSMISSION} with an arbitrary number of communication paths but a single message per \ac{UE}.
\end{proof}

Note that the non-cooperative capacity region from Theorem~\ref{t:BASICS_ZERO_CAPREGION} implicitly makes use of \ac{IRC}, as~\eqref{e:BASICS_ZERO_CAPREGION_GIVENP} exploits the spatial structure of interference. We will later also observe the performance of \ac{FDM}, where the \ac{UE}s focus their transmit power on orthogonal resources, and hence interference is avoided. As such schemes play a minor role in the context of \ac{CoMP}, however, corresponding capacity expressions are omitted.

\subsection{Base Station Cooperation Schemes For Finite Backhaul}
\label{s:BASICS_COOPSCHEMES}

We investigate four \ac{BS} cooperation schemes, constrained to scenarios with $M=K=2$ for clarity. The schemes are initially considered with only one phase of information exchange between \ac{BS}s, but the benefit of iterative \ac{BS} cooperation will be discussed in Section~\ref{s:PRACTICAL_PARALLELS}.
 
\subsubsection{Distributed Interference Subtraction (DIS)~\cite{Marsch_WCNC08}}
\label{s:BASICS_COOPSCHEMES_DIS}

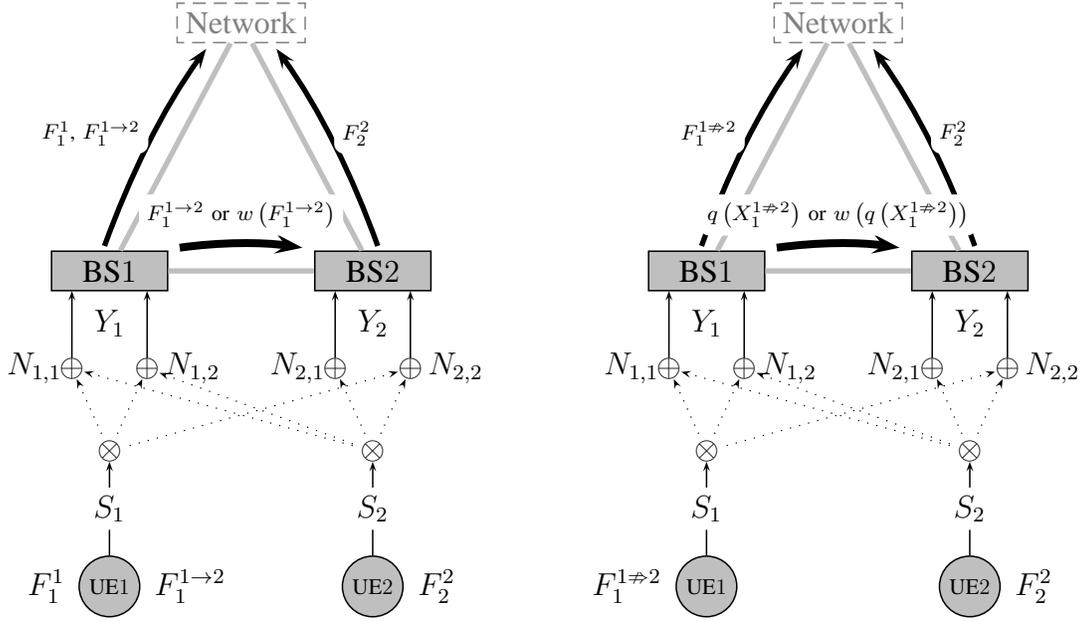
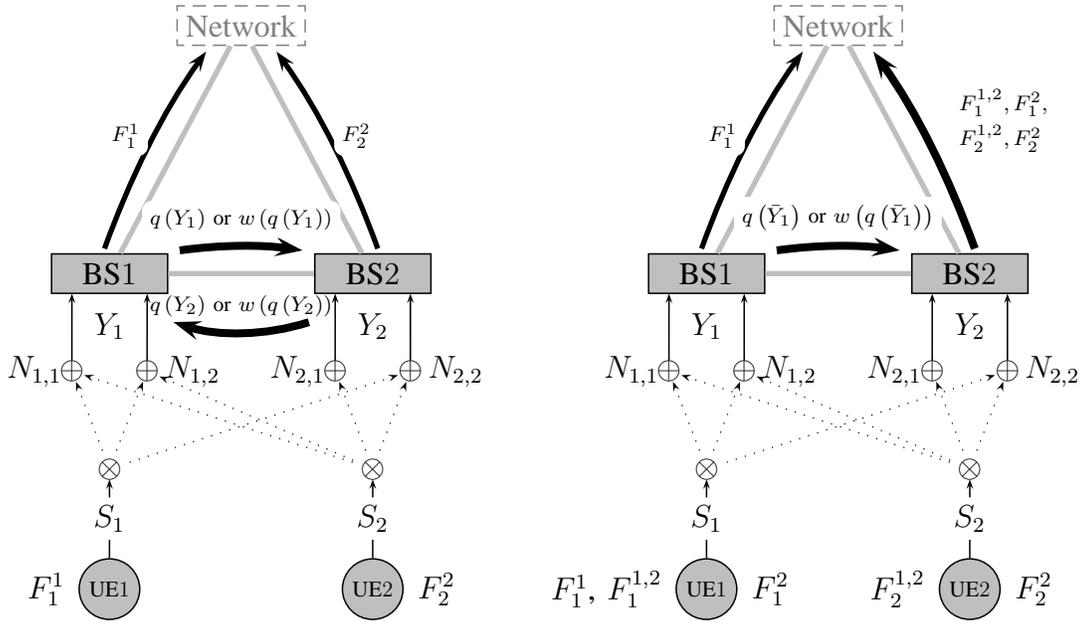
\begin{figure*}
\centerline{\subfigure[\textbf{DIS}: One BS forwards a decoded message to the other BS for (partial) interference subtraction.]{\begingroup
\unitlength=1mm
\begin{picture}(70, 90)(0, 0)
  %\put(0, 0){\framebox(88, 55)}

  \psset{xunit=1mm, yunit=1mm, linewidth=0.2mm}
  
  % Place the network
  \rput(37.5, 80){\rnode{NW}{\psframebox[linestyle=dashed, linecolor=gray]{\gray Network}}}

  % Place the base stations
  \rput(15, 47){\rnode{BS11}{\psframebox[linestyle=none]{\white B}}}
  \rput(25, 47){\rnode{BS12}{\psframebox[linestyle=none]{\white B}}}
  \rput(20, 47){\rnode{BS1}{\psframebox[fillstyle=solid, fillcolor=lightgray]{~~BS$1$~~}}}
  \rput(50, 47){\rnode{BS21}{\psframebox[linestyle=none]{\white B}}}
  \rput(60, 47){\rnode{BS22}{\psframebox[linestyle=none]{\white B}}}
  \rput(55, 47){\rnode{BS2}{\psframebox[fillstyle=solid, fillcolor=lightgray]{~~BS$2$~~}}}
    
  % Place the addition signs
  \rput(15, 34){\rnode{plus11}{$\oplus$}}
  \rput(25, 34){\rnode{plus12}{$\oplus$}}
  \rput(50, 34){\rnode{plus21}{$\oplus$}}
  \rput(60, 34){\rnode{plus22}{$\oplus$}}
  
	% place the noise terms
  \rput(10, 34){\rnode{n11}{$N_{1,1}$}}
  \rput(31, 34){\rnode{n12}{$N_{1,2}$}}
  \rput(45, 34){\rnode{n21}{$N_{2,1}$}}
  \rput(66, 34){\rnode{n22}{$N_{2,2}$}}
        
  % Place the multiplication signs
  \rput(20, 23){\rnode{mult1}{$\otimes$}}
  \rput(55, 23){\rnode{mult2}{$\otimes$}}
	
  % Place the terminals
  \rput(20, 5){\cnodeput[fillstyle=solid, fillcolor=lightgray](0, 0){UE1}{\scalebox{0.7}{UE$1$}}}
	\rput(55, 5){\cnodeput[fillstyle=solid, fillcolor=lightgray](0, 0){UE2}{\scalebox{0.7}{UE$2$}}}
	  
	% place the messages created at the terminals
	\rput[r](14, 5){$\Ful_1^1$}
  \rput[l](26, 5){$\Ful_1^{1 \rightarrow 2}$}
  \rput[l](61, 5){$\Ful_2^2$}
  
  % multiplied channels
  \ncline[linestyle=solid]{->}{h11}{mult11}
  \ncline[linestyle=solid]{->}{h12}{mult12}
  \ncline[linestyle=solid]{->}{h21}{mult21}
  \ncline[linestyle=solid]{->}{h22}{mult22}
  
  % draw connections
  \ncline[linestyle=dotted]{->}{mult1}{plus11}
  \ncline[linestyle=dotted]{->}{mult1}{plus12}
  \ncline[linestyle=dotted]{->}{mult11}{plus21}
  \ncline[linestyle=dotted]{->}{mult1}{plus22}
  \ncline[linestyle=dotted]{->}{mult2}{plus11}
  \ncline[linestyle=dotted]{->}{mult2}{plus12}
  \ncline[linestyle=dotted]{->}{mult2}{plus21}
  \ncline[linestyle=dotted]{->}{mult2}{plus22}
    
  % signals from UEs into channel
  \ncline[linestyle=solid]{->}{UE1}{mult1}\mput*{$S_1$}
  \ncline[linestyle=solid]{->}{UE2}{mult2}\mput*{$S_2$}
  
  % signals going into BSs
  \ncline[linestyle=solid]{->}{plus11}{BS11}
  \ncline[linestyle=solid]{->}{plus12}{BS12}
  \ncline[linestyle=solid]{->}{plus21}{BS21}
  \ncline[linestyle=solid]{->}{plus22}{BS22}
    
 	% place the received signals
  \rput(20, 40){\rnode{n1}{$Y_1$}}
  \rput(55, 40){\rnode{n1}{$Y_2$}}
  
  % backhaul connections
  \ncline[linestyle=solid, linecolor=lightgray, linewidth=2pt]{-}{NW}{BS1}
	\ncline[linestyle=solid, linecolor=lightgray, linewidth=2pt]{-}{NW}{BS2}
	\ncline[linestyle=solid, linecolor=lightgray, linewidth=2pt]{-}{BS1}{BS2}
		\ncarc[linestyle=solid, linewidth=2pt, offset=5pt, nodesepB=5pt]{->}{BS1}{NW}\Aput*[0.5pt, framearc=0.5]{\scriptsize $\Ful_1^1$, $\Ful_1^{1 \rightarrow 2}$}
				\ncarc[linestyle=solid, linewidth=2pt, offset=5pt, nodesepA=5pt]{<-}{NW}{BS2}\Aput*[0.5pt, framearc=0.5]{\scriptsize $\Ful_2^2$}

	\ncarc[linestyle=solid, linewidth=3pt, offset=5pt, nodesepA=5pt, nodesepB=5pt]{->}{BS1}{BS2}\Aput*[3pt, framearc=0.5]{\scriptsize $\Ful_1^{1 \rightarrow 2}$ or $w\left(\Ful_1^{1 \rightarrow 2}\right)$}
	      
  \end{picture}
\endgroup\label{f:ANALYSIS_COOPSCHEMES_DIS}} \hfil
\subfigure[\textbf{CIF}: One BS forwards quantized transmit sequences to the other BS for (partial) interference subtraction.]{\begingroup
\unitlength=1mm
\begin{picture}(70, 90)(0, 0)
  %\put(0, 0){\framebox(88, 55)}

  \psset{xunit=1mm, yunit=1mm, linewidth=0.2mm}
  
  % Place the network
  \rput(37.5, 80){\rnode{NW}{\psframebox[linestyle=dashed, linecolor=gray]{\gray Network}}}

  % Place the base stations
  \rput(15, 47){\rnode{BS11}{\psframebox[linestyle=none]{\white B}}}
  \rput(25, 47){\rnode{BS12}{\psframebox[linestyle=none]{\white B}}}
  \rput(20, 47){\rnode{BS1}{\psframebox[fillstyle=solid, fillcolor=lightgray]{~~BS$1$~~}}}
  \rput(50, 47){\rnode{BS21}{\psframebox[linestyle=none]{\white B}}}
  \rput(60, 47){\rnode{BS22}{\psframebox[linestyle=none]{\white B}}}
  \rput(55, 47){\rnode{BS2}{\psframebox[fillstyle=solid, fillcolor=lightgray]{~~BS$2$~~}}}
  
  % Place the addition signs
  \rput(15, 34){\rnode{plus11}{$\oplus$}}
  \rput(25, 34){\rnode{plus12}{$\oplus$}}
  \rput(50, 34){\rnode{plus21}{$\oplus$}}
  \rput(60, 34){\rnode{plus22}{$\oplus$}}
  
	% place the noise terms
  \rput(10, 34){\rnode{n11}{$N_{1,1}$}}
  \rput(31, 34){\rnode{n12}{$N_{1,2}$}}
  \rput(45, 34){\rnode{n21}{$N_{2,1}$}}
  \rput(66, 34){\rnode{n22}{$N_{2,2}$}}
        
  % Place the multiplication signs
  \rput(20, 23){\rnode{mult1}{$\otimes$}}
  \rput(55, 23){\rnode{mult2}{$\otimes$}}
	
  % Place the terminals
  \rput(20, 5){\cnodeput[fillstyle=solid, fillcolor=lightgray](0, 0){UE1}{\scalebox{0.7}{UE$1$}}}
	\rput(55, 5){\cnodeput[fillstyle=solid, fillcolor=lightgray](0, 0){UE2}{\scalebox{0.7}{UE$2$}}}
	  
	% place the message created at the terminals
  \rput[r](14, 5){$\Ful_1^{1 \nRightarrow 2}$}
  \rput[l](61, 5){$\Ful_2^2$}
  
  % multiplied channels
  \ncline[linestyle=solid]{->}{h11}{mult11}
  \ncline[linestyle=solid]{->}{h12}{mult12}
  \ncline[linestyle=solid]{->}{h21}{mult21}
  \ncline[linestyle=solid]{->}{h22}{mult22}
  
  % draw connections
  \ncline[linestyle=dotted]{->}{mult1}{plus11}
  \ncline[linestyle=dotted]{->}{mult1}{plus12}
  \ncline[linestyle=dotted]{->}{mult11}{plus21}
  \ncline[linestyle=dotted]{->}{mult1}{plus22}
  \ncline[linestyle=dotted]{->}{mult2}{plus11}
  \ncline[linestyle=dotted]{->}{mult2}{plus12}
  \ncline[linestyle=dotted]{->}{mult2}{plus21}
  \ncline[linestyle=dotted]{->}{mult2}{plus22}
    
  % signals from UEs into channel
  \ncline[linestyle=solid]{->}{UE1}{mult1}\mput*{$S_1$}
  \ncline[linestyle=solid]{->}{UE2}{mult2}\mput*{$S_2$}
  
  % signals going into BSs
  \ncline[linestyle=solid]{->}{plus11}{BS11}
  \ncline[linestyle=solid]{->}{plus12}{BS12}
  \ncline[linestyle=solid]{->}{plus21}{BS21}
  \ncline[linestyle=solid]{->}{plus22}{BS22}
    
 	% place the received signals
  \rput(20, 40){\rnode{n1}{$Y_1$}}
  \rput(55, 40){\rnode{n1}{$Y_2$}}
  
  % backhaul connections
  \ncline[linestyle=solid, linecolor=lightgray, linewidth=2pt]{-}{NW}{BS1}
	\ncline[linestyle=solid, linecolor=lightgray, linewidth=2pt]{-}{NW}{BS2}
	\ncline[linestyle=solid, linecolor=lightgray, linewidth=2pt]{-}{BS1}{BS2}
		\ncarc[linestyle=solid, linewidth=2pt, offset=5pt, nodesepB=5pt]{->}{BS1}{NW}\Aput*[0.5pt, framearc=0.5]{\scriptsize $\Ful_1^{1 \nRightarrow 2}$}
				\ncarc[linestyle=solid, linewidth=2pt, offset=5pt, nodesepA=5pt]{<-}{NW}{BS2}\Aput*[0.5pt, framearc=0.5]{\scriptsize $\Ful_2^2$}

	\ncarc[linestyle=solid, linewidth=3pt, offset=5pt, nodesepA=5pt, nodesepB=5pt]{->}{BS1}{BS2}\Aput*[3pt, framearc=0.5]{\scriptsize $q\left(\Xul_1^{1 \nRightarrow 2}\right)$ or $w\left(q\left(\Xul_1^{1 \nRightarrow 2}\right)\right)$}
	      
  \end{picture}
\endgroup
\label{f:ANALYSIS_COOPSCHEMES_CIF}}}
\centerline{\subfigure[\textbf{DAS-D}: Both BSs simult. exchange quantized receive signals, but decode UEs locally.]{\begingroup
\unitlength=1mm
\begin{picture}(70, 90)(0, 0)
  %\put(0, 0){\framebox(88, 55)}

  \psset{xunit=1mm, yunit=1mm, linewidth=0.2mm}
  
  % Place the network
  \rput(37.5, 80){\rnode{NW}{\psframebox[linestyle=dashed, linecolor=gray]{\gray Network}}}

  % Place the base stations
  \rput(15, 47){\rnode{BS11}{\psframebox[linestyle=none]{\white B}}}
  \rput(25, 47){\rnode{BS12}{\psframebox[linestyle=none]{\white B}}}
  \rput(20, 47){\rnode{BS1}{\psframebox[fillstyle=solid, fillcolor=lightgray]{~~BS$1$~~}}}
  \rput(50, 47){\rnode{BS21}{\psframebox[linestyle=none]{\white B}}}
  \rput(60, 47){\rnode{BS22}{\psframebox[linestyle=none]{\white B}}}
  \rput(55, 47){\rnode{BS2}{\psframebox[fillstyle=solid, fillcolor=lightgray]{~~BS$2$~~}}}
  
  % Place the addition signs
  \rput(15, 34){\rnode{plus11}{$\oplus$}}
  \rput(25, 34){\rnode{plus12}{$\oplus$}}
  \rput(50, 34){\rnode{plus21}{$\oplus$}}
  \rput(60, 34){\rnode{plus22}{$\oplus$}}
  
	% place the noise terms
  \rput(10, 34){\rnode{n11}{$N_{1,1}$}}
  \rput(31, 34){\rnode{n12}{$N_{1,2}$}}
  \rput(45, 34){\rnode{n21}{$N_{2,1}$}}
  \rput(66, 34){\rnode{n22}{$N_{2,2}$}}
        
  % Place the multiplication signs
  \rput(20, 21){\rnode{mult1}{$\otimes$}}
  \rput(55, 21){\rnode{mult2}{$\otimes$}}
	
  % Place the terminals
  \rput(20, 5){\cnodeput[fillstyle=solid, fillcolor=lightgray](0, 0){UE1}{\scalebox{0.7}{UE$1$}}}
	\rput(55, 5){\cnodeput[fillstyle=solid, fillcolor=lightgray](0, 0){UE2}{\scalebox{0.7}{UE$2$}}}
	  
	% place the message created at the terminals
  \rput[r](14, 5){$\Ful_1^1$}
  \rput[l](61, 5){$\Ful_2^2$}
  
  % multiplied channels
  \ncline[linestyle=solid]{->}{h11}{mult11}
  \ncline[linestyle=solid]{->}{h12}{mult12}
  \ncline[linestyle=solid]{->}{h21}{mult21}
  \ncline[linestyle=solid]{->}{h22}{mult22}
  
  % draw connections
  \ncline[linestyle=dotted]{->}{mult1}{plus11}
  \ncline[linestyle=dotted]{->}{mult1}{plus12}
  \ncline[linestyle=dotted]{->}{mult11}{plus21}
  \ncline[linestyle=dotted]{->}{mult1}{plus22}
  \ncline[linestyle=dotted]{->}{mult2}{plus11}
  \ncline[linestyle=dotted]{->}{mult2}{plus12}
  \ncline[linestyle=dotted]{->}{mult2}{plus21}
  \ncline[linestyle=dotted]{->}{mult2}{plus22}
    
  % signals from UEs into channel
  \ncline[linestyle=solid]{->}{UE1}{mult1}\mput*{$S_1$}
  \ncline[linestyle=solid]{->}{UE2}{mult2}\mput*{$S_2$}
  
  % signals going into BSs
  \ncline[linestyle=solid]{->}{plus11}{BS11}
  \ncline[linestyle=solid]{->}{plus12}{BS12}
  \ncline[linestyle=solid]{->}{plus21}{BS21}
  \ncline[linestyle=solid]{->}{plus22}{BS22}
    
 	% place the received signals
  \rput(20, 40){\rnode{n1}{$Y_1$}}
  \rput(55, 40){\rnode{n1}{$Y_2$}}
  
  % backhaul connections
  \ncline[linestyle=solid, linecolor=lightgray, linewidth=2pt]{-}{NW}{BS1}
	\ncline[linestyle=solid, linecolor=lightgray, linewidth=2pt]{-}{NW}{BS2}
	\ncline[linestyle=solid, linecolor=lightgray, linewidth=2pt]{-}{BS1}{BS2}
		\ncarc[linestyle=solid, linewidth=2pt, offset=5pt, nodesepB=5pt]{->}{BS1}{NW}\Aput*[0.5pt, framearc=0.5]{\scriptsize $\Ful_1^1$}
				\ncarc[linestyle=solid, linewidth=2pt, offset=5pt, nodesepA=5pt]{<-}{NW}{BS2}\Aput*[0.5pt, framearc=0.5]{\scriptsize $\Ful_2^2$}
	\ncarc[linestyle=solid, linewidth=3pt, offset=5pt, nodesepA=5pt, nodesepB=5pt]{->}{BS1}{BS2}\Aput*[3pt, framearc=0.5]{\scriptsize $q\left(Y_1\right)$ or $w\left(q\left(Y_1\right)\right)$}
	\ncarc[linestyle=solid, linewidth=3pt, offset=-10pt, nodesepA=5pt, nodesepB=5pt, arcangle=-18]{<-}{BS1}{BS2}\Aput[3pt, framearc=0.5]{\scriptsize $q\left(Y_2\right)$ or $w\left(q\left(Y_2\right)\right)$}
		      
  \end{picture}
\endgroup\label{f:ANALYSIS_COOPSCHEMES_DASD}} \hfil
\subfigure[\textbf{DAS-C}: One BS forwards quant. receive signals to the other BS for joint UE decoding.]{\begingroup
\unitlength=1mm
\begin{picture}(70, 90)(0, 0)
  %\put(0, 0){\framebox(88, 55)}

  \psset{xunit=1mm, yunit=1mm, linewidth=0.2mm}
  
  % Place the network
  \rput(37.5, 80){\rnode{NW}{\psframebox[linestyle=dashed, linecolor=gray]{\gray Network}}}

  % Place the base stations
  \rput(15, 47){\rnode{BS11}{\psframebox[linestyle=none]{\white B}}}
  \rput(25, 47){\rnode{BS12}{\psframebox[linestyle=none]{\white B}}}
  \rput(20, 47){\rnode{BS1}{\psframebox[fillstyle=solid, fillcolor=lightgray]{~~BS$1$~~}}}
  \rput(50, 47){\rnode{BS21}{\psframebox[linestyle=none]{\white B}}}
  \rput(60, 47){\rnode{BS22}{\psframebox[linestyle=none]{\white B}}}
  \rput(55, 47){\rnode{BS2}{\psframebox[fillstyle=solid, fillcolor=lightgray]{~~BS$2$~~}}}
  
  % Place the addition signs
  \rput(15, 34){\rnode{plus11}{$\oplus$}}
  \rput(25, 34){\rnode{plus12}{$\oplus$}}
  \rput(50, 34){\rnode{plus21}{$\oplus$}}
  \rput(60, 34){\rnode{plus22}{$\oplus$}}
  
	% place the noise terms
  \rput(10, 34){\rnode{n11}{$N_{1,1}$}}
  \rput(31, 34){\rnode{n12}{$N_{1,2}$}}
  \rput(45, 34){\rnode{n21}{$N_{2,1}$}}
  \rput(66, 34){\rnode{n22}{$N_{2,2}$}}
        
  % Place the multiplication signs
  \rput(20, 21){\rnode{mult1}{$\otimes$}}
  \rput(55, 21){\rnode{mult2}{$\otimes$}}
	
  % Place the terminals
  \rput(20, 5){\cnodeput[fillstyle=solid, fillcolor=lightgray](0, 0){UE1}{\scalebox{0.7}{UE$1$}}}
	\rput(55, 5){\cnodeput[fillstyle=solid, fillcolor=lightgray](0, 0){UE2}{\scalebox{0.7}{UE$2$}}}
	  
	% place the message created at the terminals
	\rput[r](14, 5){$\Ful_1^1$, $\Ful_1^{1,2}$}
  \rput[l](26, 5){$\Ful_1^2$}
  \rput[r](49, 5){$\Ful_2^{1,2}$}
  \rput[l](61, 5){$\Ful_2^2$}
  
  % multiplied channels
  \ncline[linestyle=solid]{->}{h11}{mult11}
  \ncline[linestyle=solid]{->}{h12}{mult12}
  \ncline[linestyle=solid]{->}{h21}{mult21}
  \ncline[linestyle=solid]{->}{h22}{mult22}
  
  % draw connections
  \ncline[linestyle=dotted]{->}{mult1}{plus11}
  \ncline[linestyle=dotted]{->}{mult1}{plus12}
  \ncline[linestyle=dotted]{->}{mult11}{plus21}
  \ncline[linestyle=dotted]{->}{mult1}{plus22}
  \ncline[linestyle=dotted]{->}{mult2}{plus11}
  \ncline[linestyle=dotted]{->}{mult2}{plus12}
  \ncline[linestyle=dotted]{->}{mult2}{plus21}
  \ncline[linestyle=dotted]{->}{mult2}{plus22}
    
  % signals from UEs into channel
  \ncline[linestyle=solid]{->}{UE1}{mult1}\mput*{$S_1$}
  \ncline[linestyle=solid]{->}{UE2}{mult2}\mput*{$S_2$}
  
  % signals going into BSs
  \ncline[linestyle=solid]{->}{plus11}{BS11}
  \ncline[linestyle=solid]{->}{plus12}{BS12}
  \ncline[linestyle=solid]{->}{plus21}{BS21}
  \ncline[linestyle=solid]{->}{plus22}{BS22}
    
 	% place the received signals
  \rput(20, 40){\rnode{n1}{$Y_1$}}
  \rput(55, 40){\rnode{n1}{$Y_2$}}
  
  % backhaul connections
  \ncline[linestyle=solid, linecolor=lightgray, linewidth=2pt]{-}{NW}{BS1}
	\ncline[linestyle=solid, linecolor=lightgray, linewidth=2pt]{-}{NW}{BS2}
	\ncline[linestyle=solid, linecolor=lightgray, linewidth=2pt]{-}{BS1}{BS2}
			\ncarc[linestyle=solid, linewidth=2pt, offset=5pt, nodesepB=5pt]{->}{BS1}{NW}\Aput*[0.5pt, framearc=0.5]{\scriptsize $\Ful_1^1$}
				\ncarc[linestyle=solid, linewidth=3pt, offset=5pt, nodesepA=5pt]{<-}{NW}{BS2}\Aput*[0.5pt, framearc=0.5]{$\scriptsize \begin{array}{l} \Ful_1^{1,2}, \Ful_1^2, \\ \Ful_2^{1,2}, \Ful_2^2 \end{array}$}
	\ncarc[linestyle=solid, linewidth=3pt, offset=5pt, nodesepA=5pt, nodesepB=5pt]{->}{BS1}{BS2}\Aput*[3pt, framearc=0.5]{\scriptsize $q\left(\bar{Y}_1\right)$ or $w\left(q\left(\bar{Y}_1\right)\right)$}
	      
  \end{picture}
\endgroup
\label{f:ANALYSIS_COOPSCHEMES_DASC}}}
\caption{Uplink CoMP schemes for $M=K=2$ analyzed in this work.}
\label{f:ANALYSIS_COOPSCHEMES}
\end{figure*}

This concept is often paralleled to {\em decode-and-forward} in relaying: One \ac{BS} decodes (part of) one \ac{UE}'s transmission and forwards the decoded data to the other \ac{BS}, in the \ac{CoMP} case for (partial) interference cancellation. For a particular \ac{BS}-\ac{UE} assignment $\Ba=[1,2]^T$ and cooperation direction $b=1$ as shown in Fig.~\ref{f:ANALYSIS_COOPSCHEMES_DIS}, let us assume \ac{UE} $1$ transmits messages $\Ful_1^{1}$ and $\Ful_1^{1 \rightarrow 2}$, mapped onto sequences $\Xul_1^{1}$ and $\Xul_1^{1 \rightarrow 2}$. Both messages are decoded by \ac{BS} $1$, after which message $\Ful_1^{1 \rightarrow 2}$ is forwarded to \ac{BS} $2$. As the signals received by both \ac{BS}s are correlated, we consider compressing the decoded bits via Slepian-Wolf source coding~\cite{SlepianWolf_IEEETRANS73} at \ac{BS} $1$, before forwarding $w(\Ful_1^{1 \rightarrow 2})$. At \ac{BS} $2$, message $\Ful_1^{1 \rightarrow 2}$ is reconstructed with $Y_2$ as side-information (if source coding was applied), and message $\Ful_2^2$ is decoded based on the interference-reduced receive signals $\tilde{Y}_2 = Y_2 - \hat{\BH}_2^{\textnormal{e}} \sqrt{\rho_{\Ful_1^{1 \rightarrow 2}}} e(\Ful_1^{1 \rightarrow 2})$.
\begin{theorem}[Inner bound on DIS capacity region] An inner bound on the capacity region of a \acs{DIS} setup (with any assignment $\Ba$ and cooperation direction $b$) under backhaul $\beta$ is given as
\begin{equation}
\label{e:BASICS_CAPREGION_DIS_GENERAL}
\RRuldis\left(\beta\right) = {\bigcup}_{\Ba, b, \SP \SPA : \SPA \BPmax - \BP\left(\SF_{\textnormal{all}}\right) \succeq 0} \RRuldis \left(\beta, \Ba, b, \SP\right),
\end{equation}

\noindent where all $\Br \in \RRuldis(\beta, \Ba=[1,2]^T, b=1, \SP)$ fulfill $\forall \SPA k \in \{1,2\} : \SPA r_k \geq 0$, $r_1 \leq \nu_{\Ful_1^1} + \nu_{\Ful_1^{1 \rightarrow 2}}$ and
\begin{eqnarray}
\label{e:BASICS_CAPREGION_DIS_GIVENP1}
\nu_{\Ful_1^1} &\!\!\leq\!\!& \logdet{\BI + \inv{\sigma^2\BI + \Phh_1 + \BHeff{1} \BP\left(\Ful_2^2\right) \BHeffH{1}} \BHeff{1} \BP\left(\Ful_1^1\right) \BHeffH{1}} \\
\label{e:BASICS_CAPREGION_DIS_GIVENP2}
\nu_{\Ful_1^{1 \rightarrow 2}} &\!\!\leq\!\!& \logdet{\BI + \inv{\sigma^2\BI + \Phh_1 + \BHeff{1} \BP\left(\left\{\Ful_1^1, \Ful_2^2\right\}\right) \BHeffH{1}} \BHeff{1} \BP\left(\Ful_1^{1 \rightarrow 2}\right) \BHeffH{1}} \\
\label{e:BASICS_CAPREGION_DIS_GIVENP3}
\nu_{\Ful_1^{1 \rightarrow 2}} &\!\!\leq\!\!& \beta + \underbrace{\logdet{\BI\!+\!\inv{\sigma^2\BI\!+\!\Phh_2\!+\!\BHeff{2} \BP\left(\left\{\Ful_1^1, \Ful_2^2\right\}\right) \BHeffH{2}} \!\!\! \BHeff{2} \BP\left(\Ful_1^{1 \rightarrow 2}\right) \BHeffH{2}}}_{=0 \textnormal{ without Slepian-Wolf source coding}} \\
\label{e:BASICS_CAPREGION_DIS_GIVENP4}
r_2 &\!\!\leq\!\!& \logdet{\BI + \inv{\sigma^2\BI + \Phh_2 + \BHeff{2} \BP\left(\Ful_1^1\right) \BHeffH{2}} \BHeff{2} \BP\left(\Ful_2^2\right) \BHeffH{2}},
\end{eqnarray}

\noindent where $\Phh_1$ and $\Phh_2$ are the submatrices of $\Phh$ in~\eqref{e:BASICS_EFFECTIVE_NOISE} connected to \acp{BS} $1$ and $2$, respectively.
\end{theorem}
\begin{proof}
The rate of of message $\Ful_1^{1 \rightarrow 2}$ is constrained on one hand in~\eqref{e:BASICS_CAPREGION_DIS_GIVENP2} as it has to be decoded by \ac{BS} $1$ (interfered by messages $\Ful_1^1$ and $\Ful_2^2$), and on the other hand in~\eqref{e:BASICS_CAPREGION_DIS_GIVENP3} by the rate of the backhaul plus the rate at which it could be decoded by \ac{BS} $2$ without cooperation. Message $\Ful_2^2$ can then be decoded free of interference from message $\Ful_1^{1 \rightarrow 2}$ (see Eq.~\eqref{e:BASICS_CAPREGION_DIS_GIVENP4}).
\end{proof}

Note that the decoding order at the forwarding \ac{BS} is important, i.e. the forwarded message has to be decoded first such that its rate is low w.r.t. the level of interference it represents.

\subsubsection{Compressed Interference Forwarding (CIF)~\cite{SimeoneShamai_IEEETRANS09, GriegerMarsch_GLOBECOM09}}
\label{s:BASICS_COOPSCHEMES_CIF}

This scheme is similar to \acs{DIS} in the way that both \ac{BS}s decode their \ac{UE} individually, while one \ac{BS} offers the other a certain extent of interference subtraction. Here, however, the \ac{BS}s exchange {\em quantized transmit sequences}. Using \acs{CIF}, the rate/backhaul operation point can be adjusted through choosing an appropriate degree of quantization, rather than using \ac{SPC}. Let us again fix the assignment $\Ba=[1,2]^T$ and cooperation direction $b=1$ as in Fig.~\ref{f:ANALYSIS_COOPSCHEMES_CIF}, and observe the case where \ac{BS} $1$ decodes message $\Ful_1^1$, calculates the originally transmitted sequence \scalebox{0.9}{$\Xul_1^1 = e(\Ful_1^1)$} and forwards a quantized version \scalebox{0.9}{$q(\Xul_1^1)$} to \ac{BS} $2$. We optionally consider that a source-encoded version \scalebox{0.9}{$w(q(\Xul_1^1))$} is forwarded, exploiting side-information at \ac{BS} $2$. The latter \ac{BS} then reconstructs \scalebox{0.9}{$q(\Xul_1^1)$} and computes an interference-reduced version of its received signals $\tilde{Y}_2 = Y_2 - \hat{\BH}_{2}^{\textnormal{e}} \sqrt{\rho_{\Ful_1^1}} q(\Xul_1^1)$, from which message $\Ful_2^2$ can be decoded.
\begin{theorem}[Inner bound on CIF capacity region] An inner bound on the capacity region of a \acs{CIF} setup (with any assignment $\Ba$ and cooperation direction $b$) under backhaul $\beta$ is given as
\begin{equation}
\label{e:BASICS_CAPREGION_CIF_GENERAL}
\RRulcif\left(\beta\right) = {\bigcup}_{\Ba, b, \SP \SPA : \SPA \BPmax - \BP\left(\SF_{\textnormal{all}}\right) \succeq 0} \RRulcif \left(\beta, \Ba, b, \SP\right),
\end{equation}

\noindent where all rates $\Br \in \RRulcif(\beta, \Ba=[1,2]^T, b=1, \SP)$ fulfill $\forall \SPA k \in \{1,2\}: \SPA r_k \geq 0$ and
\begin{eqnarray}
\label{e:BASICS_CAPREGION_CIF_GIVENP1}
r_1 \leq \logdet{\BI + \inv{\sigma^2\BI + \Phh_1 + \BHeff{1} \BP\left(\Ful_2^2 \right) \BHeffH{1}} \BHeff{1} \BP \left( \Ful_1^1 \right) \BHeffH{1}} \\
\label{e:BASICS_CAPREGION_CIF_GIVENP2}
r_2 \leq \logdet{\BI + \inv{\sigma^2\BI + \Phh_2 + \Bheff{2,1} \xi_{\Ful_1^1} \BheffH{2,1}} \BHeff{2} \BP \left( \Ful_2^2 \right) \BHeffH{2}} \\
\label{e:BASICS_CAPREGION_CIF_GIVENP3}
\underbrace{\xi_{\Ful_1^1} \geq \frac{\rho_{\Ful_1^1}}{\MAX(2^{\beta-2},1)}}_{\textnormal{Pract. quantizer}} \SPA \SPA \textnormal{or} \SPA \SPA \underbrace{\xi_{\Ful_1^1} \geq \frac{\rho_{\Ful_1^1}}{2^{\beta}}}_{\textnormal{rate-distortion theory}} \SPA \SPA \textnormal{or} \SPA \SPA \underbrace{\xi_{\Ful_1^1} \geq \frac {\rho_{\Ful_1^1} \cdot \kappa} {2^{\beta} - 1 + \kappa}}_{\textnormal{rate-dist. th. and source coding}} \\
\label{e:BASICS_CAPREGION_CIF_GIVENP4}
\textnormal{with} \SPA \kappa = E\left\{\left.x_1^1\left(x_1^1\right)^H\right|Y_2\right\} = \inv{1 + \rho_{\Ful_1^1} \BheffH{2,1} \inv{ \sigma^2\BI + \Phh_2 + \Bheff{2} \BP\left(\Ful_2^2 \right) \BheffH{2}} \Bheff{2,1}},
\end{eqnarray}
\noindent and where $\Bheff{m,k} \in \mathbb{C}^{[\Nbs \times 1]}$ is the effective channel between \ac{BS} $m$ and \ac{UE} $k$.
\end{theorem}
\begin{proof}
Equation~\eqref{e:BASICS_CAPREGION_CIF_GIVENP1} bounds the achievable rates of message $\Ful_1^1$ interfered by message $\Ful_2^2$, while~\eqref{e:BASICS_CAPREGION_CIF_GIVENP2} bounds the rate of message $\Ful_2^2$ that is subject to a residual extent of interference from message $\Ful_1^1$, depending on the quantization noise power $\xi_{\Ful_1^1} \in \mathbb{R}_0^+$. Equation~\eqref{e:BASICS_CAPREGION_CIF_GIVENP3} gives exactly this quantity as a function of backhaul $\beta$, where we distinguish between the cases of a practical quantizer as given in~\cite{LindeGray_IEEETRANS80} or operation on the rate-distortion bound~\cite{CoverThomas_BOOK06}, without or with source coding~\cite{WynerZiv_IEEETRANS76}. For the latter case,~\eqref{e:BASICS_CAPREGION_CIF_GIVENP4} denotes the variance of the symbols in $X_1^1$ {\em conditioned} on the signals $Y_2$ received by BS $2$.
\end{proof}

\subsubsection{Distributed Antenna System - Decentralized Decoding (DAS-D)~\cite{MarschFettweis_PIMRC08}}
\label{s:BASICS_COOPSCHEMES_DASD}

In a third cooperation scheme based on decentralized decoding, the \ac{BS}s exchange {\em quantized receive signals} rather than decoded bits or transmit sequences. For a particular assignment $\Ba=[1,2]^T$ as shown in Fig.~\ref{f:ANALYSIS_COOPSCHEMES_DASD}, we assume the \acp{UE} transmit messages $\Ful_1^{1}$ and $\Ful_2^{2}$, respectively, mapped onto sequences $\Xul_1^{1}$ and $\Xul_2^{2}$. Both \ac{BS}s now create quantized versions $q(Y_1)$, $q(Y_2)$ of their received signals, and forward these over the backhaul. Optionally, source coding can be applied, such that $w(q(Y_1))$, $w(q(Y_2))$ are exchanged. Both \ac{BS}s then use this information and their received signals to reconstruct $q(Y_1)$, $q(Y_2)$, and then decode messages $\Ful_1^1$, $\Ful_2^2$, respectively.
\begin{theorem}[Inner bound on DAS-D capacity region] An inner bound on the capacity region of \acs{DASD} (for any \ac{BS}-\ac{UE} assignment $\Ba$) under sum backhaul $\beta$ is given as
\begin{equation}
\label{e:BASICS_CAPREGION_DASD_GENERAL}
\RRuldasd\left(\beta\right) = {\bigcup}_{\Ba, \SP \SPA : \SPA \BPmax - \BP\left(\SF_{\textnormal{all}}\right) \succeq 0} \RRuldasd \left(\beta, \Ba, \SP\right),
\end{equation}

\noindent where all rates $\Br \in \RRuldasd(\beta, \Ba=[1,2]^T, \SP)$ fulfill $\forall \SPA k \in \{1,2\}: \SPA r_k \geq 0$ and
\begin{eqnarray}
\label{e:BASICS_CAPREGION_DASD_GIVENP1}
r_1 &\!\!\leq\!\!& \logdet{\BI + \inv{\BHeff{} \BP\left( \Ful_2^2 \right) \BHeffH{} + \!\! \left[\scriptsize \begin{array}{cc} \Bnull & \Bnull \\ \Bnull & \Pqq_2 \end{array}\right] + \Phh + \sigma^2\BI} \BHeff{} \BP\left( \Ful_1^1 \right) \BHeffH{}} \\
\label{e:BASICS_CAPREGION_DASD_GIVENP2}
r_2 &\!\!\leq\!\!& \logdet{\BI + \inv{\BHeff{} \BP\left( \Ful_1^1 \right) \BHeffH{} + \!\! \left[ \scriptsize \begin{array}{cc} \Pqq_1 & \Bnull \\ \Bnull & \Bnull \end{array} \right] + \Phh + \sigma^2\BI} \BHeff{} \BP\left( \Ful_2^2 \right) \BHeffH{}},
\end{eqnarray}
\begin{equation}
\label{e:BASICS_CAPREGION_DASD_BHCONSTRAINT}
\textnormal{with} \SPA \underbrace{\sum\limits_{m=1}^2 \logdet{\BI + \inv{\Pqq_m} \dg\left(\Pyy_{m}\right)} + 2M\Nbs \leq \beta}_{\textnormal{Performance of a practical quantizer}} \SPA \textnormal{or} \SPA  \underbrace{\sum\limits_{m=1}^2 \logdet{\BI + \inv{\Pqq_m} \mathbf{\Psi_m}} \leq \beta}_{\textnormal{Rate-dist. th. (opt. source coding)}}
\end{equation}

\noindent where $\mathbf{\Psi}_m$ is either the receive signal covariance at BS $m$, i.e. $\mathbf{\Psi}_m := \Pyy_m = \BHeff{m} \BP(\SFall) \BHeffH{m} + \Phh_m + \sigma^2\BI$, or the receive signal covariance {\em conditioned} on the signals received by the other \ac{BS} (if we consider source coding), i.e. $\mathbf{\Psi}_m := \Pyy_{m|m'}$ with $\forall \SPA m \in \{1,2\}, m' \neq m:$~\cite{CosoSimoens, delCosoSimoens_ISIT08}
\begin{equation}
\Pyy_{m|m'} = \BHeff{m} \inv{\BI + \BP\left( \SFall \right) \BHeffH{m'} \inv{\Phh_{m'} + \sigma^2\BI} \BHeff{m'}} \BP\left( \SFall \right) \BHeffH{m} + \Phh_m + \sigma^2\BI
\end{equation}
\end{theorem}
\begin{proof}
The message rates are constrained in~\eqref{e:BASICS_CAPREGION_DASD_GIVENP1} and~\eqref{e:BASICS_CAPREGION_DASD_GIVENP2} due to interference and quantization noise on the antennas of the corresponding remote \ac{BS}. The quantization noise covariances are limited through~\eqref{e:BASICS_CAPREGION_DASD_BHCONSTRAINT}, where we again consider a practical scheme quantizing each dimension separately (losing one bit to the rate-distortion bound in each real dimension)~\cite{LindeGray_IEEETRANS80}, operation on the rate-distortion bound~\cite{CoverThomas_BOOK06}, or the latter including source coding~\cite{CosoSimoens,delCosoSimoens_ISIT08}. 
\end{proof}

Investing different portions of backhaul into the two cooperation directions allows trading the rate of one \ac{UE} against the other. The calculation of (weighted sum-rate) optimal quantization noise covariances for~\eqref{e:BASICS_CAPREGION_DASD_BHCONSTRAINT} has been studied in detail in~\cite{CosoSimoens,delCosoSimoens_ISIT08}.

\subsubsection{Distributed Antenna System - Centralized Decoding (DAS-C)}
\label{s:BASICS_COOPSCHEMES_DASC}

We finally consider the case that both \ac{UE}s are decoded jointly by one of the \ac{BS}s, and the other \ac{BS} is degraded to a \ac{RRH} that quantizes and forwards received signals, possibly being {\em oblivious} to transmitted codewords~\cite{SanderovichShamai_IEEETRANS09}. To incorporate various proposed concepts, let us assume for a fixed $\Ba=[1,2]^T$ and $b=1$ that \ac{UE} $1$ transmits message $\Ful_1^1$, which is decoded non-cooperatively by \ac{BS} $1$. The \ac{UE}s further transmit messages $\Ful_1^{\{1,2\}}$ and $\Ful_2^{\{1,2\}}$, respectively, which are individually decoded by both \ac{BS}s, and messages $\Ful_1^2$ and $\Ful_2^2$, respectively, which are jointly decoded by \ac{BS} $2$. This model (see Fig.~\ref{f:ANALYSIS_COOPSCHEMES_DASC}), hence reflects the concept of {\em common messages}, known to be beneficial in the context of an \ac{IC}~\cite{HanKobayashi_IEEETRANS81} and for centralized detection~\cite{Marsch_WPMC08}, and also the concept of local, non-cooperative decoding~\cite{SanderovichShamai_ISIT07, SanderovichShamai_IEEETRANS07_subm, SanderovichShamai_IEEETRANS09}. \ac{BS} $1$ decodes messages $\Ful_1^1$, $\Ful_1^{\{1,2\}}$ and $\Ful_2^{\{1,2\}}$, and subtracts the corresponding transmit sequences from the received signals to construct 
\begin{equation}
\bar{Y}_1 = Y_1 - \hat{\Bh}_{1,1}^{\textnormal{e}} \left( \sqrt{\rho_{\Ful_1^{\{1,2\}}}} \Xul_1^{\{1,2\}} + \sqrt{\rho_{\Ful_1^1}} \Xul_1^1 \right) - \hat{\Bh}_{1,2}^{\textnormal{e}} \cdot \sqrt{ \rho_{\Ful_2^{\{1,2\}}}} \cdot \Xul_2^{\{1,2\}}.
\end{equation}

This is then quantized to $q(\bar{Y}_1)$ (and optionally source-encoded to $w(q(\bar{Y}_1))$) and forwarded to \ac{BS} $2$. The latter \ac{BS} also decodes messages \scalebox{0.9}{$\Ful_1^{\{1,2\}}$} and \scalebox{0.9}{$\Ful_2^{\{1,2\}}$}, subtracts their impact on $Y_2$, and uses $Y_2$ plus the information provided by \ac{BS} $1$ to finally decode messages $\Ful_1^2$ and $\Ful_2^2$.
\begin{theorem}[Inner bound on DAS-C capacity region] An inner bound on the capacity region of \acs{DASC} (for any assignment $\Ba$ and cooperation direction $b$) under backhaul $\beta$ is given as
\begin{equation}
\label{e:BASICS_CAPREGION_DASC_GENERAL}
\RRuldasc\left(\beta\right) = {\bigcup}_{\SP \SPA : \SPA \BPmax - \BP\left(\SF_{\textnormal{all}}\right) \succeq 0} \RRuldasc \left(\beta, \Ba, b, \SP \right),
\end{equation}

\noindent where all rate tuples $\Br \in \RRuldasc(\beta, \Ba=[1,2]^T, b=1, \SP)$ fulfill $\forall \SPA k \in \{1,2\}: \SPA r_k \geq 0$ and
\begin{eqnarray}
\label{e:BASICS_CAPREGION_DASC_GIVENP1}
r_1 &\!\!\!=\!\!\!& \nu_{\Ful_1^1} + \nu_{\Ful_1^{1,2}} + \nu_{\Ful_1^2} \SPA \textnormal{and} \SPA r_2 = \nu_{\Ful_2^{1,2}} + \nu_{\Ful_2^2} \\
\label{e:BASICS_CAPREGION_DASC_GIVENP2}
\forall \SPA \SF \subseteq \left\{\Ful_1^1, \Ful_1^{1,2}, \Ful_2^{1,2} \right\} : \sum\limits_{F \in \SF} \nu_{F} &\!\!\!\leq\!\!\!& \logdet{\BI + \inv{\BHeff{1} \BP\left(\left\{\Ful_1^1, \Ful_1^2, \Ful_2^2\right\}\right) \BHeffH{1}\!\!+\!\!\Pnn_1} \!\!\!\!\! \mathbf{\Psi}_1} \\
\label{e:BASICS_CAPREGION_DASC_GIVENP3}
\forall \SPA \SF \subseteq \left\{\Ful_1^{1,2}, \Ful_2^{1,2} \right\} : \sum\limits_{F \in \SF} \nu_{F} &\!\!\!\leq\!\!\!& \logdet{\BI + \inv{\BHeff{2} \BP\left(\left\{\Ful_1^2, \Ful_2^2 \right\}\right) \BHeffH{2}\!\!+\!\!\Pnn_2} \!\!\!\!\! \mathbf{\Psi}_2} \\
\label{e:BASICS_CAPREGION_DASC_GIVENP4}
\forall \SPA \SF \subseteq \left\{\Ful_1^2, \Ful_2^2\right\} : \sum\limits_{F \in \SF} \nu_{F} &\!\!\!\leq\!\!\!& \logdet{\BI + \inv{\arr{cc}{\Pqq & \Bnull \\ \Bnull & \rho_{\Ful_1^1}\Bheff{2,1}\BheffH{2,1}}\!\!+\!\!\Pnn} \!\!\!\!\! \mathbf{\Psi}},
\end{eqnarray}

\noindent where $\Pnn = \Phh + \sigma^2\BI$ and $\mathbf{\Psi}=\BHeff{} \BP(\SP) \BHeffH{}$, and we have the backhaul constraint
\begin{equation}
\label{e:BASICS_CAPREGION_DASC_BHCONSTRAINT}
\underbrace{\LOG_2 \left| \BI\!\!+\!\!\inv{\Pqq}\!\!\dg\left(\barPyy_{1}\right) \right| + 2\Nbs \leq \beta}_{\textnormal{Pract. quant.}}, \SPA \SPA \underbrace{\LOG_2 \left| \BI\!\!+\!\!\inv{\Pqq}\!\!\barPyy_{1} \right| \leq \beta}_{\textnormal{}} \SPA \SPA \textnormal{or} \SPA \SPA \underbrace{\LOG_2 \left| \BI\!\!+\!\!\inv{\Pqq}\!\!\barPyy_{1|2} \right| \leq \beta}_{\textnormal{Source coding}},
\end{equation}
\noindent where $\barPyy_{1}$ is the signal covariance at \ac{BS} $1$ after the subtraction of decoded messages, and $\barPyy_{1|2}$ is the same quantity, but conditioned on the signals at \ac{BS} $2$ after message subtraction.
\end{theorem}
\begin{proof}
Eq.~\eqref{e:BASICS_CAPREGION_DASC_GIVENP1} states that the overall \ac{UE} rates are the sum of the rates of the superimposed messages. Eqs.~\eqref{e:BASICS_CAPREGION_DASC_GIVENP2} and~\eqref{e:BASICS_CAPREGION_DASC_GIVENP3} state the sum rates of any tuples of messages decoded without \ac{BS} cooperation by \acp{BS} $1$ and $2$, respectively, and~\eqref{e:BASICS_CAPREGION_DASC_GIVENP4} states the sum rate bound on the two messages $\Ful_1^2$ and $\Ful_2^2$ that are jointly decoded by \ac{BS} $2$. In the latter equation, we have to consider not only quantization noise (making the same differentiations w.r.t. quantization as before), but also the fact that the signals received at \ac{BS} $2$ are still subject to interference from message $\Ful_1^1$, as this message is not known to \ac{BS} $2$. The backhaul constraint in~\eqref{e:BASICS_CAPREGION_DASC_BHCONSTRAINT} is based on~\cite{CosoSimoens,delCosoSimoens_ISIT08}. 
\end{proof}

In this work, we also consider \ac{FDM} scenarios where the \ac{UE}s are served on orthogonal resources, but enhanced through the exchange of received signals over the backhaul. We will see that these schemes play a minor role, and hence omit equations for brevity.

\subsection{Performance Regions}
\label{s:BASICS_PERFREGIONS}

The concept of {\em performance regions} was introduced in~\cite{Marsch_WCNC08} to jointly capture achievable rate tuples and the corresponding backhaul requirement. A {\em performance point} is defined as
\begin{equation}
\label{s:BASICS_PERFPOINTS}
Z = \left\langle \Br, \beta \right\rangle,
\end{equation}

\noindent and a {\em performance region} connected to an arbitrary \ac{BS} cooperation scheme {\em yz} is defined as
\begin{equation}
\label{e:BASICS_PERFREGIONS}
\SZ^{\textnormal{yz}} = \bigcup \left\{ \left\langle \Br, \beta \right\rangle : \SPA \Br \in \RRyz \left(\beta \right) \right\}.
\end{equation}

\begin{figure}
\centerline{\input{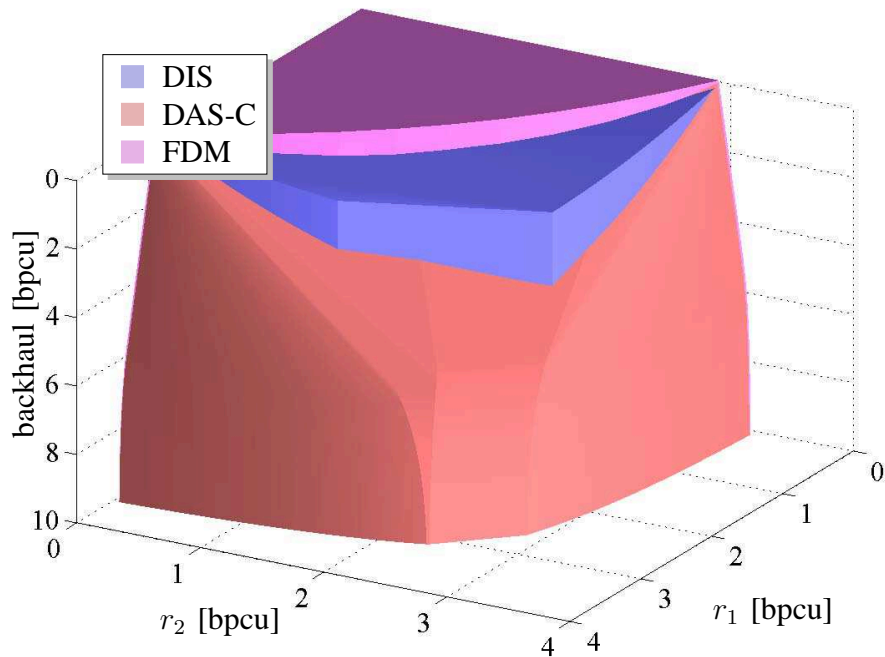}}
\caption{Illustration of a performance region for an example channel with $M=K=2$ and $\Nbs=1$.}
\label{f:BASICS_PERFREGION}
\end{figure}

Note that the convex hull operation $\bigcup$ in~\eqref{e:BASICS_PERFREGIONS} implies the option of time-sharing along the backhaul dimension, while each region $\RRyz$ already incorporates time-sharing between different \ac{BS}-\ac{UE} assignments and cooperation directions. An example performance region is shown in Fig.~\ref{f:BASICS_PERFREGION} for \acs{DIS}, \acs{CIF}, \acs{DASD}, \acs{DASC} or \acs{FDM} (all assuming practical quantization and no~\ac{SPC}) for $M=K=2$, $\Nbs=1$ and $\BH = [1, \sqrt{0.25}; \sqrt{0.5}, 1]$. We observe imperfect \ac{CSI} with $\Npilots=2$, and set $\sigma^2 = 0.1$ (\ac{SISO} \ac{SNR} of $10$ dB on the main links). We plot the achievable \ac{UE} rates on the x- and y-axis, and the required backhaul on the z-axis. The top surface of the performance region hence reflects the capacity region in the non-cooperative case, while its intersection with the x-y plane inner bounds the capacity region for infinite \ac{BS} cooperation. Note that the latter deviates from a pentagon shape~\cite{CoverThomas_BOOK06} due to imperfect \ac{CSI}. For the example channel, \acs{FDM} schemes are beneficial in the regime of no or very limited backhaul, \acs{DIS} concepts are interesting for moderate backhaul, whereas \acs{DASC} is the only scheme approaching \ac{MAC} performance for large backhaul. \acs{DASD} and \acs{CIF} are inferior for all extents of backhaul and hence not visible.

\section{Analysis of Cooperation Concepts}
\label{s:ANALYSIS}

\subsection{Scenarios and Channels Considered}
\label{s:ANALYSIS_CHANNELS}

In the sequel, we are interested in the sum-rate achievable in a small scenario with $M=K=2$ and $\Nbs=2$, for infinite or no \ac{BS} cooperation, and for the \ac{CoMP} schemes from Section~\ref{s:BASICS_COOPSCHEMES}. We observe different scenarios characterized by the location of the \ac{UE}s, where for each \ac{UE} $k$ a normalized distance $d_k \in [0,1]$ denotes whether it is close to its assigned \ac{BS} (small $d_k$), at the cell-edge ($d_k=0.5$), or closer to the other \ac{BS} ($d_k > 0.5$). We use exemplary channel matrices
\begin{equation}
\label{e:ANALYSIS_CHANNEL_M2K2}
\BH = \left[ \begin{array}{cc}\sqrt{\lambda_{1,1}} & \sqrt{\lambda_{1,2}}e^{j(-\varphi_{1}/2-\varphi_{12}/2)} \\
\sqrt{\lambda_{1,1}} & \sqrt{\lambda_{1,2}}e^{j(+\varphi_{1}/2-\varphi_{12}/2)} \\
\sqrt{\lambda_{2,1}}e^{j(-\varphi_{2}/2-\varphi_{12}/2)} & \sqrt{\lambda_{2,2}} \\
\sqrt{\lambda_{2,1}}e^{j(+\varphi_{2}/2-\varphi_{12}/2)} & \sqrt{\lambda_{2,2}} \end{array} \right] \SPA \SPA \textnormal{with}
\end{equation}
\begin{equation}
\label{e:ANALYSIS_PATHGAIN_M2K2}
\forall k \in \{1,2\}, \SPA m \neq k : \SPA \lambda_{k,k} = \frac{d_k^{-\theta}}{d_k^{-\theta} + \left(1-d_k\right)^{-\theta}}, \SPA \SPA \lambda_{m,k} = \frac{\left(1-d_k\right)^{-\theta}}{d_k^{-\theta} + \left(1-d_k\right)^{-\theta}},
\end{equation}

\noindent where $\lambda_{m,k}$ is the linear path gain from \ac{UE} $k$ to \ac{BS} $m$, based on a flat-plane pathloss model with pathloss exponent $\theta=3.5$, is normalized by the transmit power of the \ac{UE}s. We here assume {\em multi-cell power control}, where the {\em average} power by which a \ac{UE} is received by both \ac{BS}s is normalized to $1$. We can then use $\BPmax=\BI$ regardless of \ac{UE} location, and the dominant links are normalized to unit gain in the case of $d_k=0$. Terms $\varphi_{1}$, $\varphi_{2}$ and $\varphi_{12}$ in~\eqref{e:ANALYSIS_CHANNEL_M2K2} are phases connected to the orthogonality of the channels seen by \ac{BS} $1$ or $2$, respectively, or the additional orthogonality of the compound channel. Unless stated otherwise, we observe channels of average orthogonality, i.e. $\varphi_{1}=\varphi_{2}=\varphi_{12}=\pi/2$, and choose $\sigma^2=0.1$, leading with~\eqref{e:ANALYSIS_PATHGAIN_M2K2} to a cell-center \ac{SISO} \ac{SNR} of $10$ dB, which is motivated through system level simulations in~\cite{Marsch_DISS10}. 

\subsection{Overall CoMP Gain under Imperfect CSI}
\label{s:ANALYSIS_COMPGAIN}

\begin{figure}
\centerline{\begingroup
\unitlength=1mm
\psset{xunit=157.50000mm, yunit=5.88889mm, linewidth=0.1mm}
\psset{arrowsize=2pt 3, arrowlength=1.4, arrowinset=.4}\psset{axesstyle=frame}
\begin{pspicture}(0.13651, 0.30189)(0.60000, 11.00000)
\psaxes[subticks=0, labels=all, xsubticks=1, ysubticks=1, Ox=0.2, Oy=2, Dx=0.1, Dy=1]{-}(0.20000, 2.00000)(0.20000, 2.00000)(0.60020, 11.00020)%
\multips(0.30000, 2.00000)(0.10000, 0.0){3}{\psline[linecolor=black, linestyle=dotted, linewidth=0.2mm](0, 0)(0, 9.00000)}
\multips(0.20000, 3.00000)(0, 1.00000){8}{\psline[linecolor=black, linestyle=dotted, linewidth=0.2mm](0, 0)(0.40000, 0)}
\rput[b](0.40000, 0.30189){$d_1 = d_2$}
\psclip{\psframe(0.20000, 2.00000)(0.60000, 11.00000)}
\psline[linecolor=black, plotstyle=curve, linestyle=solid, dotscale=1.5 1.5, showpoints=false, linewidth=0.3mm](0.20000, 7.93596)(0.22000, 7.84326)(0.24000, 7.71979)(0.26000, 7.56214)(0.28000, 7.37009)(0.30000, 7.14740)(0.32000, 6.90120)(0.34000, 6.63985)(0.36000, 6.37034)(0.38000, 6.09652)(0.40000, 5.81856)(0.42000, 5.53360)(0.44000, 5.23693)(0.46000, 4.92336)(0.48000, 4.86565)(0.50000, 4.88252)(0.52000, 4.86565)(0.54000, 4.92336)(0.56000, 5.23693)(0.58000, 5.53360)(0.60000, 5.81856)
\psline[linecolor=black, plotstyle=curve, dotstyle=square, linestyle=none, dotscale=1.5 1.5, showpoints=true, linewidth=0.3mm](0.20000, 7.93596)(0.25000, 7.64097)(0.30000, 7.14740)(0.35000, 6.50510)(0.40000, 5.81856)(0.45000, 5.08014)(0.50000, 4.88252)(0.55000, 5.08014)(0.60000, 5.81856)
\psline[linecolor=black, plotstyle=curve, linestyle=solid, dotscale=1.5 1.5, showpoints=false, linewidth=0.3mm](0.20000, 7.96313)(0.22000, 7.89924)(0.24000, 7.82504)(0.26000, 7.74364)(0.28000, 7.65866)(0.30000, 7.57336)(0.32000, 7.48992)(0.34000, 7.40934)(0.36000, 7.33176)(0.38000, 7.25708)(0.40000, 7.18569)(0.42000, 7.11908)(0.44000, 7.06014)(0.46000, 7.01303)(0.48000, 6.98231)(0.50000, 6.97160)(0.52000, 6.98231)(0.54000, 7.01303)(0.56000, 7.06014)(0.58000, 7.11908)(0.60000, 7.18569)
\psline[linecolor=black, plotstyle=curve, dotstyle=o, linestyle=none, dotscale=1.5 1.5, showpoints=true, linewidth=0.3mm](0.20000, 7.96313)(0.25000, 7.78434)(0.30000, 7.57336)(0.35000, 7.37055)(0.40000, 7.18569)(0.45000, 7.03659)(0.50000, 6.97160)(0.55000, 7.03659)(0.60000, 7.18569)
\psline[linecolor=darkgray, plotstyle=curve, linestyle=dotted, dotscale=1.5 1.5, showpoints=false, linewidth=0.3mm](0.20000, 9.06159)(0.22000, 8.95313)(0.24000, 8.81980)(0.26000, 8.66387)(0.28000, 8.48975)(0.30000, 8.30287)(0.32000, 8.10822)(0.34000, 7.90893)(0.36000, 7.70552)(0.38000, 7.49587)(0.40000, 7.27564)(0.42000, 7.03915)(0.44000, 6.78023)(0.46000, 6.53803)(0.48000, 6.59967)(0.50000, 6.62061)(0.52000, 6.59967)(0.54000, 6.53803)(0.56000, 6.78023)(0.58000, 7.03915)(0.60000, 7.27564)
\psline[linecolor=darkgray, plotstyle=curve, dotstyle=square, linestyle=none, dotscale=1.5 1.5, showpoints=true, linewidth=0.3mm](0.20000, 9.06159)(0.25000, 8.74184)(0.30000, 8.30287)(0.35000, 7.80723)(0.40000, 7.27564)(0.45000, 6.65913)(0.50000, 6.62061)(0.55000, 6.65913)(0.60000, 7.27564)
\psline[linecolor=darkgray, plotstyle=curve, linestyle=dotted, dotscale=1.5 1.5, showpoints=false, linewidth=0.3mm](0.20000, 9.30037)(0.22000, 9.29508)(0.24000, 9.28773)(0.26000, 9.27776)(0.28000, 9.26451)(0.30000, 9.24727)(0.32000, 9.22532)(0.34000, 9.19811)(0.36000, 9.16534)(0.38000, 9.12732)(0.40000, 9.08523)(0.42000, 9.04145)(0.44000, 8.99965)(0.46000, 8.96451)(0.48000, 8.94089)(0.50000, 8.93254)(0.52000, 8.94089)(0.54000, 8.96451)(0.56000, 8.99965)(0.58000, 9.04145)(0.60000, 9.08523)
\psline[linecolor=darkgray, plotstyle=curve, dotstyle=o, linestyle=none, dotscale=1.5 1.5, showpoints=true, linewidth=0.3mm](0.20000, 9.30037)(0.25000, 9.28275)(0.30000, 9.24727)(0.35000, 9.18172)(0.40000, 9.08523)(0.45000, 8.98208)(0.50000, 8.93254)(0.55000, 8.98208)(0.60000, 9.08523)
\psline[linecolor=darkgray, plotstyle=curve, linestyle=dashed, dotscale=1.5 1.5, showpoints=false, linewidth=0.3mm](0.20000, 7.16760)(0.22000, 7.09354)(0.24000, 6.99323)(0.26000, 6.86174)(0.28000, 6.69541)(0.30000, 6.49317)(0.32000, 6.25753)(0.34000, 5.99434)(0.36000, 5.71126)(0.38000, 5.41540)(0.40000, 5.11141)(0.42000, 4.80074)(0.44000, 4.48209)(0.46000, 4.15270)(0.48000, 3.98291)(0.50000, 3.99567)(0.52000, 3.98291)(0.54000, 4.15270)(0.56000, 4.48209)(0.58000, 4.80074)(0.60000, 5.11141)
\psline[linecolor=darkgray, plotstyle=curve, dotstyle=square, linestyle=none, dotscale=1.5 1.5, showpoints=true, linewidth=0.3mm](0.20000, 7.16760)(0.25000, 6.92749)(0.30000, 6.49317)(0.35000, 5.85280)(0.40000, 5.11141)(0.45000, 4.31739)(0.50000, 3.99567)(0.55000, 4.31739)(0.60000, 5.11141)
\psline[linecolor=darkgray, plotstyle=curve, linestyle=dashed, dotscale=1.5 1.5, showpoints=false, linewidth=0.3mm](0.20000, 7.17917)(0.22000, 7.11846)(0.24000, 7.04276)(0.26000, 6.95291)(0.28000, 6.85128)(0.30000, 6.74144)(0.32000, 6.62737)(0.34000, 6.51280)(0.36000, 6.40074)(0.38000, 6.29348)(0.40000, 6.19309)(0.42000, 6.10205)(0.44000, 6.02379)(0.46000, 5.96272)(0.48000, 5.92355)(0.50000, 5.91002)(0.52000, 5.92355)(0.54000, 5.96272)(0.56000, 6.02379)(0.58000, 6.10205)(0.60000, 6.19309)
\psline[linecolor=darkgray, plotstyle=curve, dotstyle=o, linestyle=none, dotscale=1.5 1.5, showpoints=true, linewidth=0.3mm](0.20000, 7.17917)(0.25000, 6.99783)(0.30000, 6.74144)(0.35000, 6.45677)(0.40000, 6.19309)(0.45000, 5.99325)(0.50000, 5.91002)(0.55000, 5.99325)(0.60000, 6.19309)
\endpsclip
\rput[t]{90}(0.13651, 6.50000){sum rate [bits/channel use]}
\pcline[linewidth=2pt]{->}(0.50000,4.90000)(0.50000,7.00000)\Aput*{+43\%}
\rput[l](0.20635, 10.49057){SISO SNR 10dB}
\rput[r](0.59365, 2.50943){sum rate max.}
\psframe[linecolor=black, fillstyle=solid, fillcolor=white, shadowcolor=lightgray, shadowsize=1mm, shadow=true](0.21270, 2.33962)(0.37143, 6.75472)
\rput[l](0.26349, 6.24528){inf. coop.}
\psline[linecolor=black, linestyle=none, linewidth=0.3mm](0.22540, 6.24528)(0.25079, 6.24528)
\psline[linecolor=black, linestyle=none, linewidth=0.3mm](0.22540, 6.24528)(0.25079, 6.24528)
\psdots[linecolor=black, linestyle=none, linewidth=0.3mm, dotstyle=o, dotscale=1.5 1.5, linecolor=black](0.23810, 6.24528)
\rput[l](0.26349, 5.39623){no coop.}
\psline[linecolor=black, linestyle=none, linewidth=0.3mm](0.22540, 5.39623)(0.25079, 5.39623)
\psline[linecolor=black, linestyle=none, linewidth=0.3mm](0.22540, 5.39623)(0.25079, 5.39623)
\psdots[linecolor=black, linestyle=none, linewidth=0.3mm, dotstyle=square, dotscale=1.5 1.5, linecolor=black](0.23810, 5.39623)
\rput[l](0.26349, 4.54717){$\Npilots=\infty$}
\psline[linecolor=darkgray, linestyle=dotted, linewidth=0.3mm](0.22540, 4.54717)(0.25079, 4.54717)
\rput[l](0.26349, 3.69811){$\Npilots=2$}
\psline[linecolor=black, linestyle=solid, linewidth=0.3mm](0.22540, 3.69811)(0.25079, 3.69811)
\rput[l](0.26349, 2.84906){$\Npilots=1$}
\psline[linecolor=darkgray, linestyle=dashed, linewidth=0.3mm](0.22540, 2.84906)(0.25079, 2.84906)

\end{pspicture}
\endgroup
 }
\caption{Overall gain through BS cooperation for $M=K=2$, $\Nbs=2$, and channels of average orthogonality.}
\label{f:ANALYSIS_ZEROINF}
\end{figure}
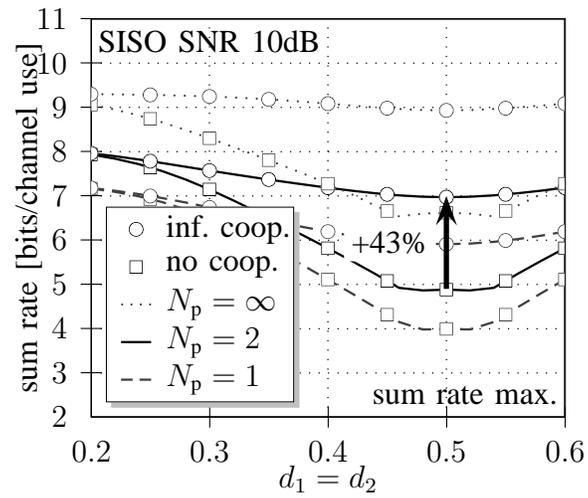

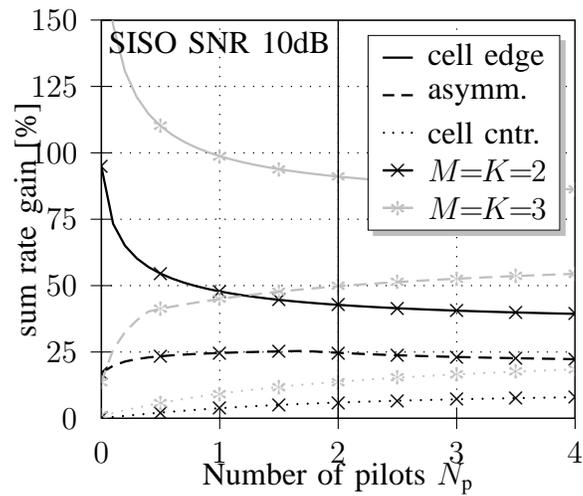
\begin{figure}
\centerline{\begingroup
\unitlength=1mm
\psset{xunit=15.75000mm, yunit=0.35333mm, linewidth=0.1mm}
\psset{arrowsize=2pt 3, arrowlength=1.4, arrowinset=.4}\psset{axesstyle=frame}
\begin{pspicture}(-0.63492, -28.30189)(4.00000, 150.00000)
\psaxes[subticks=0, labels=all, xsubticks=1, ysubticks=1, Ox=0, Oy=0, Dx=1, Dy=25]{-}(0.00000, 0.00000)(0.00000, 0.00000)(4.00020, 150.00020)%
\multips(1.00000, 0.00000)(1.00000, 0.0){3}{\psline[linecolor=black, linestyle=dotted, linewidth=0.2mm](0, 0)(0, 150.00000)}
\multips(0.00000, 25.00000)(0, 25.00000){5}{\psline[linecolor=black, linestyle=dotted, linewidth=0.2mm](0, 0)(4.00000, 0)}
\rput[b](2.00000, -28.30189){Number of pilots $\Np$}
\psclip{\psframe(0.00000, 0.00000)(4.00000, 150.00000)}
\psline[linecolor=black, plotstyle=curve, linestyle=solid, dotscale=1.5 1.5, showpoints=false, linewidth=0.3mm](0.00000, 95.04925)(0.10000, 73.28476)(0.20000, 65.05788)(0.30000, 60.25314)(0.40000, 56.96740)(0.50000, 54.52946)(0.60000, 52.62735)(0.70000, 51.09139)(0.80000, 49.81944)(0.90000, 48.74549)(1.00000, 47.82459)(1.10000, 47.02488)(1.20000, 46.32303)(1.30000, 45.70149)(1.40000, 45.14680)(1.50000, 44.64839)(1.60000, 44.19789)(1.70000, 43.78853)(1.80000, 43.41478)(1.90000, 43.07209)(2.00000, 42.75666)(2.10000, 42.46529)(2.20000, 42.19528)(2.30000, 41.94433)(2.40000, 41.71045)(2.50000, 41.49192)(2.60000, 41.28727)(2.70000, 41.09518)(2.80000, 40.91453)(2.90000, 40.74431)(3.00000, 40.58362)(3.10000, 40.43168)(3.20000, 40.28779)(3.30000, 40.15131)(3.40000, 40.02168)(3.50000, 39.89840)(3.60000, 39.78100)(3.70000, 39.66906)(3.80000, 39.56222)(3.90000, 39.46012)(4.00000, 39.36246)
\psline[linecolor=black, plotstyle=curve, dotstyle=x, linestyle=none, dotscale=1.5 1.5, showpoints=true, linewidth=0.3mm](0.00000, 95.04925)(0.50000, 54.52946)(1.00000, 47.82459)(1.50000, 44.64839)(2.00000, 42.75666)(2.50000, 41.49192)(3.00000, 40.58362)(3.50000, 39.89840)(4.00000, 39.36246)
\psline[linecolor=black, plotstyle=curve, linestyle=dashed, dotscale=1.5 1.5, showpoints=false, linewidth=0.3mm](0.00000, 17.11957)(0.10000, 19.81059)(0.20000, 21.40823)(0.30000, 22.34769)(0.40000, 22.96749)(0.50000, 23.41468)(0.60000, 23.75865)(0.70000, 24.03547)(0.80000, 24.26556)(0.90000, 24.46134)(1.00000, 24.63087)(1.10000, 24.77964)(1.20000, 24.91155)(1.30000, 25.02950)(1.40000, 25.13568)(1.50000, 25.23184)(1.60000, 25.31933)(1.70000, 25.39928)(1.80000, 25.19048)(1.90000, 24.93424)(2.00000, 24.69998)(2.10000, 24.48505)(2.20000, 24.28721)(2.30000, 24.10453)(2.40000, 23.93539)(2.50000, 23.77836)(2.60000, 23.63223)(2.70000, 23.49592)(2.80000, 23.36851)(2.90000, 23.24916)(3.00000, 23.13717)(3.10000, 23.03188)(3.20000, 22.93272)(3.30000, 22.83920)(3.40000, 22.75085)(3.50000, 22.66728)(3.60000, 22.58810)(3.70000, 22.51301)(3.80000, 22.44168)(3.90000, 22.37387)(4.00000, 22.30931)
\psline[linecolor=black, plotstyle=curve, dotstyle=x, linestyle=none, dotscale=1.5 1.5, showpoints=true, linewidth=0.3mm](0.00000, 17.11957)(0.50000, 23.41468)(1.00000, 24.63087)(1.50000, 25.23184)(2.00000, 24.69998)(2.50000, 23.77836)(3.00000, 23.13717)(3.50000, 22.66728)(4.00000, 22.30931)
\psline[linecolor=black, plotstyle=curve, linestyle=dotted, dotscale=1.5 1.5, showpoints=false, linewidth=0.3mm](0.00000, 0.27363)(0.10000, 0.56558)(0.20000, 0.95690)(0.30000, 1.36915)(0.40000, 1.77853)(0.50000, 2.17420)(0.60000, 2.55116)(0.70000, 2.90742)(0.80000, 3.24259)(0.90000, 3.55717)(1.00000, 3.85209)(1.10000, 4.12851)(1.20000, 4.38767)(1.30000, 4.63081)(1.40000, 4.85912)(1.50000, 5.07373)(1.60000, 5.27568)(1.70000, 5.46596)(1.80000, 5.64546)(1.90000, 5.81499)(2.00000, 5.97531)(2.10000, 6.12710)(2.20000, 6.27097)(2.30000, 6.40751)(2.40000, 6.53724)(2.50000, 6.66062)(2.60000, 6.77809)(2.70000, 6.89005)(2.80000, 6.99687)(2.90000, 7.09887)(3.00000, 7.19637)(3.10000, 7.28966)(3.20000, 7.37898)(3.30000, 7.46458)(3.40000, 7.54668)(3.50000, 7.62549)(3.60000, 7.70120)(3.70000, 7.77397)(3.80000, 7.84398)(3.90000, 7.91138)(4.00000, 7.97631)
\psline[linecolor=black, plotstyle=curve, dotstyle=x, linestyle=none, dotscale=1.5 1.5, showpoints=true, linewidth=0.3mm](0.00000, 0.27363)(0.50000, 2.17420)(1.00000, 3.85209)(1.50000, 5.07373)(2.00000, 5.97531)(2.50000, 6.66062)(3.00000, 7.19637)(3.50000, 7.62549)(4.00000, 7.97631)
\psline[linecolor=lightgray, plotstyle=curve, linestyle=solid, dotscale=1.5 1.5, showpoints=false, linewidth=0.3mm](0.00000, 191.52657)(0.10000, 148.18899)(0.20000, 130.71861)(0.30000, 120.95926)(0.40000, 114.62408)(0.50000, 110.13463)(0.60000, 106.76404)(0.70000, 104.12770)(0.80000, 102.00169)(0.90000, 100.24613)(1.00000, 98.76882)(1.10000, 97.50632)(1.20000, 96.41345)(1.30000, 95.45711)(1.40000, 94.61242)(1.50000, 93.86032)(1.60000, 93.18592)(1.70000, 92.57743)(1.80000, 92.02536)(1.90000, 91.52201)(2.00000, 91.06102)(2.10000, 90.63712)(2.20000, 90.24591)(2.30000, 89.88364)(2.40000, 89.54716)(2.50000, 89.23372)(2.60000, 88.94100)(2.70000, 88.66696)(2.80000, 88.40983)(2.90000, 88.16807)(3.00000, 87.94030)(3.10000, 87.72532)(3.20000, 87.52207)(3.30000, 87.32959)(3.40000, 87.14703)(3.50000, 86.97364)(3.60000, 86.80873)(3.70000, 86.65168)(3.80000, 86.50193)(3.90000, 86.35898)(4.00000, 86.22237)
\psline[linecolor=lightgray, plotstyle=curve, dotstyle=asterisk, linestyle=none, dotscale=1.5 1.5, showpoints=true, linewidth=0.3mm](0.00000, 191.52657)(0.50000, 110.13463)(1.00000, 98.76882)(1.50000, 93.86032)(2.00000, 91.06102)(2.50000, 89.23372)(3.00000, 87.94030)(3.50000, 86.97364)(4.00000, 86.22237)
\psline[linecolor=lightgray, plotstyle=curve, linestyle=dashed, dotscale=1.5 1.5, showpoints=false, linewidth=0.3mm](0.00000, 13.56988)(0.10000, 25.85521)(0.20000, 32.46058)(0.30000, 36.94260)(0.40000, 40.27298)(0.50000, 41.30152)(0.60000, 42.10841)(0.70000, 42.88187)(0.80000, 43.61696)(0.90000, 44.31203)(1.00000, 44.96740)(1.10000, 45.58450)(1.20000, 46.16529)(1.30000, 46.71201)(1.40000, 47.22695)(1.50000, 47.71236)(1.60000, 48.17038)(1.70000, 48.60303)(1.80000, 49.01218)(1.90000, 49.39957)(2.00000, 49.76677)(2.10000, 50.11526)(2.20000, 50.44636)(2.30000, 50.76130)(2.40000, 51.06120)(2.50000, 51.34707)(2.60000, 51.61986)(2.70000, 51.88042)(2.80000, 52.12955)(2.90000, 52.36797)(3.00000, 52.59634)(3.10000, 52.81528)(3.20000, 53.02536)(3.30000, 53.22709)(3.40000, 53.42096)(3.50000, 53.60742)(3.60000, 53.78688)(3.70000, 53.95973)(3.80000, 54.12632)(3.90000, 54.28699)(4.00000, 54.44203)
\psline[linecolor=lightgray, plotstyle=curve, dotstyle=asterisk, linestyle=none, dotscale=1.5 1.5, showpoints=true, linewidth=0.3mm](0.00000, 13.56988)(0.50000, 41.30152)(1.00000, 44.96740)(1.50000, 47.71236)(2.00000, 49.76677)(2.50000, 51.34707)(3.00000, 52.59634)(3.50000, 53.60742)(4.00000, 54.44203)
\psline[linecolor=lightgray, plotstyle=curve, linestyle=dotted, dotscale=1.5 1.5, showpoints=false, linewidth=0.3mm](0.00000, 1.21388)(0.10000, 2.06990)(0.20000, 3.06771)(0.30000, 4.03825)(0.40000, 4.95441)(0.50000, 5.81119)(0.60000, 6.61023)(0.70000, 7.35538)(0.80000, 8.05105)(0.90000, 8.70161)(1.00000, 9.31112)(1.10000, 9.88330)(1.20000, 10.42147)(1.30000, 10.92860)(1.40000, 11.40735)(1.50000, 11.86008)(1.60000, 12.28890)(1.70000, 12.69571)(1.80000, 13.08220)(1.90000, 13.44990)(2.00000, 13.80017)(2.10000, 14.13427)(2.20000, 14.45330)(2.30000, 14.75830)(2.40000, 15.05020)(2.50000, 15.32983)(2.60000, 15.59798)(2.70000, 15.85534)(2.80000, 16.10259)(2.90000, 16.34030)(3.00000, 16.56904)(3.10000, 16.78931)(3.20000, 17.00159)(3.30000, 17.20630)(3.40000, 17.40386)(3.50000, 17.59464)(3.60000, 17.77898)(3.70000, 17.95722)(3.80000, 18.12965)(3.90000, 18.29657)(4.00000, 18.45822)
\psline[linecolor=lightgray, plotstyle=curve, dotstyle=asterisk, linestyle=none, dotscale=1.5 1.5, showpoints=true, linewidth=0.3mm](0.00000, 1.21388)(0.50000, 5.81119)(1.00000, 9.31112)(1.50000, 11.86008)(2.00000, 13.80017)(2.50000, 15.32983)(3.00000, 16.56904)(3.50000, 17.59464)(4.00000, 18.45822)
\pcline[linewidth=0.5pt]{-}(2,0)(2,150)
\endpsclip
\rput[t]{90}(-0.73492, 75.00000){sum rate gain [\%]}
\rput[l](0.06349, 141.50943){SISO SNR 10dB}
\psframe[linecolor=black, fillstyle=solid, fillcolor=white, shadowcolor=lightgray, shadowsize=1mm, shadow=true](2.24921, 70.75472)(3.87302, 144.33962)
\rput[l](2.75714, 135.84906){cell edge}
\psline[linecolor=black, linestyle=solid, linewidth=0.3mm](2.37619, 135.84906)(2.63016, 135.84906)
\rput[l](2.75714, 121.69811){asymm.}
\psline[linecolor=black, linestyle=dashed, linewidth=0.3mm](2.37619, 121.69811)(2.63016, 121.69811)
\rput[l](2.75714, 107.54717){cell cntr.}
\psline[linecolor=black, linestyle=dotted, linewidth=0.3mm](2.37619, 107.54717)(2.63016, 107.54717)
\rput[l](2.75714, 93.39623){$M\!\!=\!\!K\!\!=\!\!2$}
\psline[linecolor=black, linestyle=solid, linewidth=0.3mm](2.37619, 93.39623)(2.63016, 93.39623)
\psline[linecolor=black, linestyle=solid, linewidth=0.3mm](2.37619, 93.39623)(2.63016, 93.39623)
\psdots[linecolor=black, linestyle=solid, linewidth=0.3mm, dotstyle=x, dotscale=1.5 1.5, linecolor=black](2.50317, 93.39623)
\rput[l](2.75714, 79.24528){$M\!\!=\!\!K\!\!=\!\!3$}
\psline[linecolor=lightgray, linestyle=solid, linewidth=0.3mm](2.37619, 79.24528)(2.63016, 79.24528)
\psline[linecolor=lightgray, linestyle=solid, linewidth=0.3mm](2.37619, 79.24528)(2.63016, 79.24528)
\psdots[linecolor=lightgray, linestyle=solid, linewidth=0.3mm, dotstyle=asterisk, dotscale=1.5 1.5, linecolor=lightgray](2.50317, 79.24528)

\end{pspicture}
\endgroup
 }
\caption{Gain of BS cooperation as a function of CSI accuracy.}
\label{f:ANALYSIS_CHNEST}
\end{figure}

Let us first observe the gain from no to infinite \ac{BS} cooperation according to the bounds from Sections~\ref{s:BASICS_INF} and~\ref{s:BASICS_ZERO}. Fig.~\ref{f:ANALYSIS_ZEROINF} shows the achievable sum-rate of both \ac{UE}s, while these are simultaneously moved from the cell-center ($d_k=0.2$) to the cell-edge and slightly beyond ($d_k=0.6$). We observe different extents of \ac{CSI}, with $\Np \in \{1,2,\infty\}$. As intuitive, the \ac{CoMP} gain is largest at the cell-edge (for average channel orthogonality), and diminishes towards the cell-center. In conjunction with multi-cell power control, \ac{CoMP} can provide fairly homogeneous performance throughout all scenarios, hence improve {\em fairness}. Interestingly, the relative gain of \ac{CoMP} at the cell-edge increases with decreasing \ac{CSI}, while the opposite is the case towards the cell-center, as shown in Fig.~\ref{f:ANALYSIS_CHNEST}. The former is due to array gain from which channel estimation can benefit, while the latter is the case as the weak interference links become difficult to estimate, and hence cannot be exploited for CoMP. Fig.~\ref{f:ANALYSIS_CHNEST} also shows results for $M=K=3$, where the \ac{CoMP} gains are larger, as (for $\Nbs=2$) each \ac{BS} by itself cannot spatially separate all $3$ \ac{UE}s.

\subsection{Performance of CoMP Schemes for Specific Channels}
\label{s:ANALYSIS_SPECIFIC}

\begin{figure}
\centerline{\input{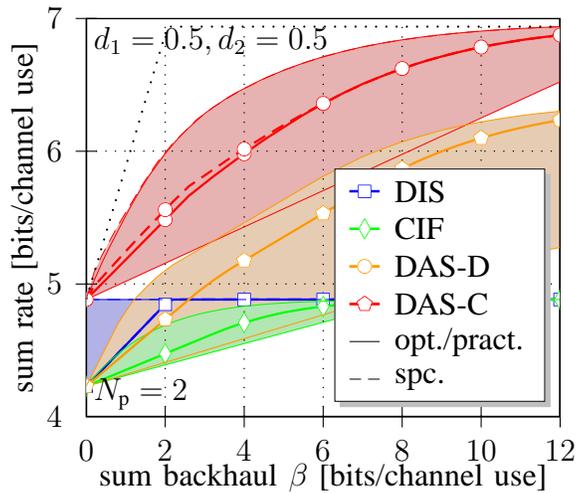}}
\caption{Sum-rate as a function of backhaul for a symmetric cell-edge scenario. For each scheme, an area indicates the range between the theoretical performance limit (employing source coding) and the performance of a practical quantizer~\cite{LindeGray_IEEETRANS80}.}
\label{f:ANALYSIS_RATEVSBH1}
\end{figure}

\begin{figure}
\centerline{\input{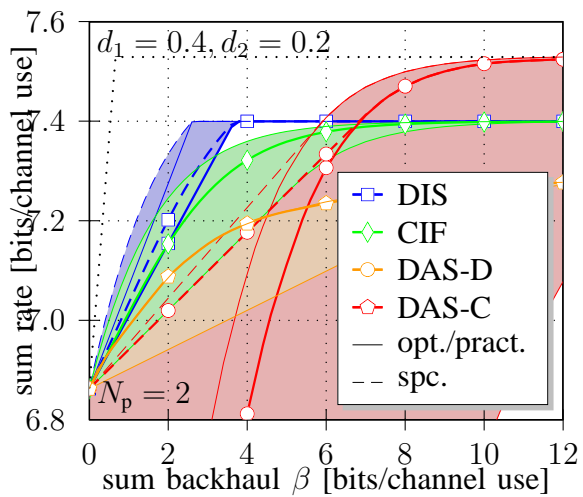}}
\caption{Sum-rate as a function of backhaul for a channel of moderate, asymmetric interference.}
\label{f:ANALYSIS_RATEVSBH2}
\end{figure}

Let us now analyze the rate/backhaul trade-off achievable with the \ac{CoMP} schemes from Section~\ref{s:BASICS_COOPSCHEMES}. We observe a scenario with $d_1=d_2=0.5$, hence a symmetric cell-edge case, in Fig.~\ref{f:ANALYSIS_RATEVSBH1}, and a scenario with $d_1=0.4$ and $d_2=0.2$, hence asymmetric and weaker interference, in Fig.~\ref{f:ANALYSIS_RATEVSBH2}. For all schemes, we show (from right to left, i.e. from less to more efficient) the performance based on a practical quantizer~\cite{LindeGray_IEEETRANS80} (not applicable to \acs{DIS}), that given by the rate-distortion bound~\cite{CoverThomas_BOOK06}, or through additional source coding. The dotted line shows the {\em cut-set bound}~\cite{CoverThomas_BOOK06}, resembling the case where each backhaul bit leads to an equal sum-rate increase of one bit, until \ac{MAC} performance is reached. In the cell-edge case, \acs{DASC} is superior for any extent of backhaul, and source coding is highly beneficial, as the correlation of received signals is strong. The gap to the cut-set bound is due to the fact that backhaul is inevitably wasted into the quantization of noise~\cite{CosoSimoens, delCosoSimoens_ISIT08}. Dashed lines indicate the (marginal) benefit of \ac{SPC}, which here can be attributed to the fact that common messages can be decoded by the \acp{BS} without cooperation, reducing the extent of signal power that is quantized for cooperation~\cite{MarschFettweis_PIMRC08}. In this symmetric cell-edge case, \acs{DIS} and \acs{CIF} yield no gain, as each \ac{BS} can decode both \ac{UE}s without cooperation. \acs{DASD} provides array gain, but is inferior to \acs{DASC} due to its inability to perform interference cancellation. In the asymmetrical scenario in Fig.~\ref{f:ANALYSIS_RATEVSBH2}, \acs{DIS} and \acs{CIF} are superior in regimes of low backhaul. Here, the cell-edge \ac{UE} $1$ is decoded first (under little interference), and then the decoded bits or a quantized transmit sequence are provided from \ac{BS} $1$ to \ac{BS} $2$. It can be shown that \acs{DIS} with \ac{SPC} is always superior to \ac{CIF}~\cite{Marsch_DISS10}, while the latter has practical advantages. While gains from source coding have decreased due to less signal correlation, it is beneficial to use \acs{DASC} with \ac{SPC}, more specifically with the option of local, non-cooperative decoding, as pointed out in~\cite{SanderovichShamai_ISIT07, ShamaiPoor_JWCC07, SanderovichShamai_IEEETRANS09}. However, the performance is inferior to that of a simple time-share between a decentralized and centralized approach~\cite{Marsch_DISS10}.

\begin{figure}
\centerline{\input{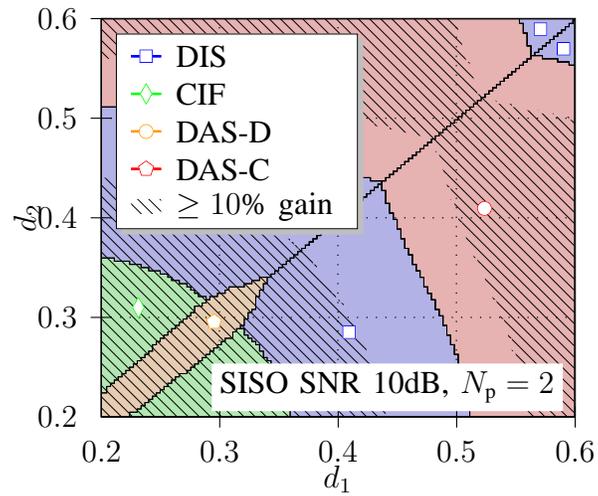}}
\caption{Best cooperation scheme as a function of UE locations, for a scenario with $M=K=2$, $\Nbs=2$ and $\beta=4$ bpcu.}
\label{f:ANALYSIS_AREAPLOT}
\end{figure}

Fig.~\ref{f:ANALYSIS_AREAPLOT} shows the best cooperation scheme as a function of $d_1$ and $d_2$, for a fixed backhaul of $\beta=4$ bits per channel use, summarizing and extending previous observations. \acs{DASC} is clearly superior in regimes of strong, possibly asymmetric interference, \acs{DIS} in regimes of weaker, asymmetric interference, \acs{CIF} in regimes of even weaker interference, while \acs{DASD} is only interesting for very weak and highly symmetric interference. Considering that the \ac{CoMP} gain in the latter regimes is marginal, it appears sufficient to adapt between \acs{DASC} and \acs{DIS}. The hashed areas in Fig.~\ref{f:ANALYSIS_AREAPLOT} indicate where such adaptation yields more than $10$\% sum-rate benefit.

\subsection{Benefit of Source Coding and Superposition Coding}
\label{s:ANALYSIS_SPCSWWZ}

Fig.~\ref{f:ANALYSIS_SPCSWWZGAIN} shows the sum-rate gain (in \%) of using source coding and/or \ac{SPC} for \acs{DIS} and \ac{DASC} (taking the maximum gain over all regimes of backhaul). As noted before, the gain of source coding techniques can be substantial for cases of strong interference, but these lead to a significantly increased complexity~\cite{XiongCheng_IEEESPM04}. A main problem as that such schemes require the interference covariance to remain constant over a reasonable extent of time, which is questionable in a cellular uplink that is typically subject to a {\em flashlight effect}, i.e. to quickly changing background interference due to scheduling. Investing effort into \ac{SPC} is clearly not attractive, as the rate increase or improvement of quantization efficiency through common messages is marginal. Further, the gain of partial local decoding is marginal for \ac{DIS}, or can be superceeded by simple time-sharing between different cooperation strategies for \ac{DASC}.

\subsection{Monte Carlo Simulation Results for $M=K=3$}
\label{s:ANALYSIS_MONTECARLO}

We now provide Monte Carlo simulation results for a scenario with $M=K=3$. Here, average path gains are generated from~\eqref{e:ANALYSIS_PATHGAIN_M2K2} for $d_1\!\!=\!\!d_2\!\!=\!\!d_3\!\!=\!\!0.5$ and $d_1\!\!=\!\!d_2\!\!=\!\!d_3\!\!=\!\!0.3$, and then many Rayleigh fading realizations are generated that fulfill $E\{|h_{2(m-1)+1,k}|^2\}=E\{|h_{2m,k}|^2\}=\lambda_{m,k}$, providing different channel orthogonalities. We compare the following schemes:
\begin{itemize}
\item Non-cooperative detection employing \ac{MRC}
\item Non-cooperative detection based on \ac{IRC}
\item Non-coop. detection with \ac{IRC} and arbitrary \ac{BS}-\ac{UE} assignment (see Section~\ref{s:BASICS_ZERO})
\item Only \ac{DIS} concepts, only \ac{DASC} concepts, or hybrid combinations, as modeled in~\cite{Marsch_WCNC08}
\item All \ac{BS}s quantize and forward to a central network entity~\cite{ShamaiPoor_JWCC07, ShamaiZaidel_PIMRC08}, denoted \acs{DASN}
\item Backhaul-enhanced \ac{FDM}, as mentioned at the end of Section~\ref{s:BASICS_COOPSCHEMES}
\end{itemize}

We again consider both information theoretical limits with or without source coding, and performance based on practical quantization. In Fig.~\ref{f:ANALYSIS_MONTECARLO1}, for the cell-edge case, we can see that \ac{IRC} is already substantially beneficial over \ac{MRC}, and an instantaneous \ac{BS}-\ac{UE} assignment further improves non-cooperative performance. For cooperation, pure \acs{DASC} strategies appear best, even under practical quantization. \acs{FDM} is strongly inferior, as the avoidance of interference is inefficient when backhaul is available, and \acs{DASN} has the disadvantage of performing quantization over one more link than schemes based on centralized decoding by a \ac{BS}. In Fig.~\ref{f:ANALYSIS_MONTECARLO2}, in the cell-center case, we can see a significant benefit of adapting between \acs{DASC} and \acs{DIS}, especially for practical quantization schemes. Under such adaptation, about $50$\% of \ac{CoMP} gain can be achieved with about $1.5$ bits of backhaul per bit of sum-rate.

\section{Practical Considerations}
\label{s:PRACTICAL}

\subsection{Parallels between Theory and Practice, and the Value of Iterative BS Cooperation}
\label{s:PRACTICAL_PARALLELS}

The previous section has revealed a central trade-off inherent to uplink \ac{CoMP}:
\begin{itemize}
\item If \ac{BS}s operate {\em code-aware}, hence perform (partial) decoding prior to cooperation, any backhaul-usage is more efficient (see results for \acs{DIS} and \acs{CIF}), but the schemes fail to achieve \ac{MAC} performance in regimes of large backhaul.
\item If \ac{BS}s are {\em oblivious} to the used codeword, backhaul is wasted into the quantization of noise, but the schemes (i.e. \acs{DASC}) asymptotically obtain the complete \ac{CoMP} gain.
\end{itemize}

Proposed practical algorithms typically perform a combination of both strategies. In, e.g., ~\cite{KhattakFettweis_VTC07, KhattakFettweis_VTC08, KhattakFettweis_EURASIP08}, each \ac{BS} (partially) decodes both the strongest interferer and its own \ac{UE}, and forwards soft-bits to the other \ac{BS}. Hence, code-awareness is used to exploit the structure in signals and interference for efficient backhaul usage, while the soft-bits inherit information on uncertainty, which yields array and diversity gain. The fact that terminal rates are strongly constrained through the first (partial) decoding process can be alleviated by using {\em iterative} \ac{BS} cooperation~\cite{BavarianCavers_GLOBECOM07, AktasHanly_IEEETRANS08, WangTse_Imprint09}, hence starting with coarse decoding and refining this in each iteration. It has been shown in~\cite{GriegerFettweis_PIMRC09, Marsch_DISS10}, however, that for the case of iterative \acs{DIS} and even under very theoretical considerations, the rate/backhaul trade-off is only marginally improved over one-shot cooperation (though the asymptotic sum-rate is improved). In practice, every backhaul usage will always inherit additional redundancy (and introduce latency), hence rendering iterative schemes even more questionable, as also observed in~\cite{MayerHagenauer_ICC06}.     

\subsection{CSI Distribution and Complexity Issues}
\label{s:PRACTICAL_CSIDISTRIBUTION_COMPLEXITY}

Table~\ref{t:PRACTICAL_SUMMARY} summarizes key aspects of the \ac{CoMP} concepts treated in this work, and adds considerations connected to the required distribution of \ac{CSI} and complexity. \acs{DIS} and \acs{CIF}, for example, have the advantage that each \ac{BS} only requires local \ac{CSI} connected to their sub-part of the channel, while \ac{DASC} requires knowledge on the compound channel at the decoding \ac{BS}, hence requiring the distribution of \ac{CSI} over the backhaul. In terms of complexity, \ac{CIF} offers the benefit that it does not require re-modulation by a \ac{BS} that performs (partial) interference subtraction. Complexity increases drastically if source coding (Wyner-Ziv, Slepian-Wolf) is performed.

\section{Conclusions}
\label{s:CONCLUSIONS}

Different theoretical uplink \ac{CoMP} concepts have been analyzed with a special focus on a constrained backhaul infrastructure and imperfect \ac{CSI}. The work has shown that strongest CoMP gains can be expected at the cell-edge, and in fact increase for diminishing \ac{CSI}, whereas gains quickly vanish towards the cell-center. This reduces the set of attractive \ac{CoMP} concepts to {\acs{DASC}}, interesting in regimes of strong interference and based on oblivious \ac{BS}s, and \acs{DIS}, based on local decoding and an exchange of decoded bits, where adaptation has shown to be beneficial. Various proposed concepts based on \ac{SPC} have shown to be of minor interest, while source coding appears attractive, but has to be put in perspective to major implementation challenges. A comparison of these theoretical concepts to proposed practical algorithms has shown the fundamental trade-off between efficient backhaul usage and maximum \ac{CoMP} gain that has to be made, and put the practical usage of iterative \ac{BS} cooperation into question.

  \appendix
  
We here sketch the proof of Theorem~\ref{t:BASICSCHNEST_EQTRANS}, providing details in~\cite{Marsch_DISS10}. Eqs.~\eqref{e:BASICS_TRANSMISSION} and~\eqref{e:BASICS_CHNEST} yield
\begin{equation}
\label{e:PROOF_TXEQ_IMPCSI_TRANSMISSION_CHANGED}
\By = \BH \Bs + \Bn = \left( \BHest - \BE \right) \Bs + \Bn = \left( \hatBHeff{} - \BEeff \right) \Bs + \Bn,
\end{equation}

\noindent where $\hatBHeff{}$ is an {\em unbiased} channel estimate, and $\BEeff$ is an uncorrelated estimation error with
\begin{equation}
\label{e:PROOF_TXEQ_IMPCSI_COVARIANCE}
\forall i,j : \SPA \hatheffij \! = \! \frac {\hat{h}_{i,j}} {\sqrt{1 \!\! + \!\! \sigma_E^2 \! \left/ \! E\left\{\left|h_{i,j}\right|^2\right\}\right.}} \SPA \textnormal{and} \SPA E\left\{|\eeffij|^2\right\} \! = \! E\left\{\left|e_{i,j}\right|^2 \left| \hat{h}_{i,j} \right. \right\} \! = \! \frac {E\left\{\left|h_{i,j}\right|^2\right\} \cdot \sigma_E^2} {E\left\{\left|h_{i,j}\right|^2\right\} \! + \! \sigma_E^2}.
\end{equation}

Treating product $\BEeff\Bs$ in~\eqref{e:PROOF_TXEQ_IMPCSI_TRANSMISSION_CHANGED} as a Gaussian random variable with a different realization in each channel access leads to an overestimation of the impact of imp. CSI~\cite{YooGoldsmith_IEEETRANS06}, i.e. we can state
\begin{equation}
\label{e:PROOF_TXEQ_IMPCSI_MUTUALINFORMATION_LOWERBOUND}
E_{\BEeff{}} \left\{ I\left( S; Y | \hatBHeff{} \right) \right\} \geq \LOG_2 \left| \BI + \inv{\Phh + \sigma^2\BI} \hatBHeff{} \BP \hatBHeffH{} \right|.
\end{equation}

In this work, we are interested in observing the rates achievable with a fixed channel $\BH$, averaged over many channel estimation realizations $\hatBHeff{}$, which we can approximate by 
\begin{equation}
\label{e:PROOF_TXEQ_IMPCSI_MUTUALINFORMATION_LOWERBOUND_FINAL}
I\left( S; Y\right) \geq E_{\hatBHeff{}} \left\{ I\left( S; Y | \hatBHeff{} \right) \right\} \approx \LOG_2 \left| \BI + \inv{\Phh + \sigma^2\BI} \BHeff{} \BP \BHeffH{} \right|.
\end{equation}

\noindent with $\BHeff{}$ given in~\eqref{e:BASICS_EFFECTIVE_CHANNEL}. Clearly, the RHS of~\eqref{e:PROOF_TXEQ_IMPCSI_MUTUALINFORMATION_LOWERBOUND_FINAL} is larger or equal to~\eqref{e:PROOF_TXEQ_IMPCSI_MUTUALINFORMATION_LOWERBOUND} due to Jensen's inequality, but numerical evaluation has shown that this aspect is negligible unless noise power and channel power are of the same order, especially in consideration of the noise overestimation in~\eqref{e:PROOF_TXEQ_IMPCSI_MUTUALINFORMATION_LOWERBOUND}.
  
\bibliographystyle{IEEEtran}
\bibliography{General_Bibliography}

%\begin{figure*}
%\centerline{\subfigure{\input{figure_ITANALYSIS_DL_M2K2_PATHLOSSMODEL}\label{f:figure_ITANALYSIS_M2K2_PATHLOSSMODEL}} \hfil
%\subfigure{\input{figure_ITANALYSIS_DL_M3K3_PATHLOSSMODEL}
%\label{f:figure_ITANALYSIS_M3K3_PATHLOSSMODEL}}}\caption{Scenarios considered in the theoretical analysis in this chapter.}
%\label{f:ANALYSIS_PATHLOSSMODEL}
%\end{figure*} 

\begin{figure}
\centerline{\begingroup
\unitlength=1mm
\psset{xunit=157.50000mm, yunit=3.53333mm, linewidth=0.1mm}
\psset{arrowsize=2pt 3, arrowlength=1.4, arrowinset=.4}\psset{axesstyle=frame}
\begin{pspicture}(0.13651, -2.83019)(0.60000, 15.00000)
\psaxes[subticks=0, labels=all, xsubticks=1, ysubticks=1, Ox=0.2, Oy=0, Dx=0.1, Dy=2]{-}(0.20000, 0.00000)(0.20000, 0.00000)(0.60020, 15.00020)%
\multips(0.30000, 0.00000)(0.10000, 0.0){3}{\psline[linecolor=black, linestyle=dotted, linewidth=0.2mm](0, 0)(0, 15.00000)}
\multips(0.20000, 2.00000)(0, 2.00000){7}{\psline[linecolor=black, linestyle=dotted, linewidth=0.2mm](0, 0)(0.40000, 0)}
\rput[b](0.40000, -2.83019){$d_1 = d_2$}
\psclip{\psframe(0.20000, 0.00000)(0.60000, 15.00000)}
\psline[linecolor=blue, plotstyle=curve, linestyle=solid, dotscale=1.5 1.5, showpoints=false, linewidth=0.1mm](0.20000, 0.00057)(0.22000, 0.00249)(0.24000, 0.00937)(0.26000, 0.02886)(0.28000, 0.07973)(0.30000, 0.18960)(0.32000, 0.39215)(0.34000, 0.75426)(0.36000, 1.30046)(0.38000, 2.05774)(0.40000, 3.16161)(0.42000, 4.48308)(0.44000, 6.35310)(0.46000, 8.33016)(0.48000, 11.37701)(0.50000, 15.43244)(0.52000, 11.37701)(0.54000, 8.33016)(0.56000, 6.35310)(0.58000, 4.48308)(0.60000, 3.16161)
\psline[linecolor=blue, plotstyle=curve, dotstyle=square, linestyle=none, dotscale=1.5 1.5, showpoints=true, linewidth=0.1mm](0.20000, 0.00057)(0.30000, 0.18960)(0.40000, 3.16161)(0.50000, 15.43244)(0.60000, 3.16161)
\psline[linecolor=blue, plotstyle=curve, linestyle=dashed, dotscale=1.5 1.5, showpoints=false, linewidth=0.3mm](0.20000, 0.04935)(0.22000, 0.10045)(0.24000, 0.18468)(0.26000, 0.30651)(0.28000, 0.45857)(0.30000, 0.61687)(0.32000, 0.74287)(0.34000, 0.79179)(0.36000, 0.72876)(0.38000, 0.53654)(0.40000, 0.23535)(0.42000, 0.00000)(0.44000, 0.00000)(0.46000, 0.00000)(0.48000, 0.00000)(0.50000, 0.00000)(0.52000, 0.00000)(0.54000, 0.00000)(0.56000, 0.00000)(0.58000, 0.00000)(0.60000, 0.23535)
\psline[linecolor=blue, plotstyle=curve, dotstyle=square, linestyle=none, dotscale=1.5 1.5, showpoints=true, linewidth=0.3mm](0.20000, 0.04935)(0.30000, 0.61687)(0.40000, 0.23535)(0.50000, 0.00000)(0.60000, 0.23535)
\psline[linecolor=blue, plotstyle=curve, linestyle=solid, dotscale=1.5 1.5, showpoints=false, linewidth=0.3mm](0.20000, 0.04977)(0.22000, 0.10224)(0.24000, 0.19134)(0.26000, 0.32775)(0.28000, 0.51672)(0.30000, 0.75685)(0.32000, 1.04392)(0.34000, 1.38003)(0.36000, 1.79026)(0.38000, 2.34459)(0.40000, 3.20906)(0.42000, 4.52303)(0.44000, 6.36342)(0.46000, 8.33016)(0.48000, 11.37701)(0.50000, 15.43244)(0.52000, 11.37701)(0.54000, 8.33016)(0.56000, 6.36342)(0.58000, 4.52303)(0.60000, 3.20906)
\psline[linecolor=blue, plotstyle=curve, dotstyle=square, linestyle=none, dotscale=1.5 1.5, showpoints=true, linewidth=0.3mm](0.20000, 0.04977)(0.30000, 0.75685)(0.40000, 3.20906)(0.50000, 15.43244)(0.60000, 3.20906)
\psline[linecolor=red, plotstyle=curve, linestyle=solid, dotscale=1.5 1.5, showpoints=false, linewidth=0.1mm](0.20000, 0.21217)(0.22000, 0.43991)(0.24000, 0.70479)(0.26000, 1.12071)(0.28000, 1.69591)(0.30000, 2.60256)(0.32000, 3.75242)(0.34000, 4.71637)(0.36000, 5.87601)(0.38000, 6.87130)(0.40000, 7.69643)(0.42000, 8.26029)(0.44000, 8.72406)(0.46000, 8.90046)(0.48000, 9.14556)(0.50000, 9.18608)(0.52000, 9.14556)(0.54000, 8.90046)(0.56000, 8.72406)(0.58000, 8.26029)(0.60000, 7.69643)
\psline[linecolor=red, plotstyle=curve, dotstyle=o, linestyle=none, dotscale=1.5 1.5, showpoints=true, linewidth=0.1mm](0.20000, 0.21217)(0.30000, 2.60256)(0.40000, 7.69643)(0.50000, 9.18608)(0.60000, 7.69643)
\psline[linecolor=red, plotstyle=curve, linestyle=dashed, dotscale=1.5 1.5, showpoints=false, linewidth=0.3mm](0.20000, 95.90047)(0.22000, 93.88296)(0.24000, 91.20544)(0.26000, 87.80855)(0.28000, 83.71386)(0.30000, 77.55919)(0.32000, 68.72808)(0.34000, 58.93140)(0.36000, 48.70173)(0.38000, 38.50180)(0.40000, 28.65113)(0.42000, 19.31368)(0.44000, 10.52354)(0.46000, 2.22399)(0.48000, 1.36205)(0.50000, 1.45650)(0.52000, 1.36205)(0.54000, 2.22399)(0.56000, 10.52354)(0.58000, 19.31368)(0.60000, 28.65113)
\psline[linecolor=red, plotstyle=curve, dotstyle=o, linestyle=none, dotscale=1.5 1.5, showpoints=true, linewidth=0.3mm](0.20000, 95.90047)(0.30000, 77.55919)(0.40000, 28.65113)(0.50000, 1.45650)(0.60000, 28.65113)
\psline[linecolor=red, plotstyle=curve, linestyle=solid, dotscale=1.5 1.5, showpoints=false, linewidth=0.3mm](0.20000, 95.90047)(0.22000, 93.88296)(0.24000, 91.20544)(0.26000, 87.80855)(0.28000, 83.71386)(0.30000, 77.55919)(0.32000, 68.72808)(0.34000, 58.93141)(0.36000, 48.70174)(0.38000, 38.50181)(0.40000, 28.65116)(0.42000, 19.31371)(0.44000, 10.57158)(0.46000, 8.90046)(0.48000, 9.10232)(0.50000, 9.32818)(0.52000, 9.10232)(0.54000, 8.90046)(0.56000, 10.57158)(0.58000, 19.31371)(0.60000, 28.65116)
\psline[linecolor=red, plotstyle=curve, dotstyle=o, linestyle=none, dotscale=1.5 1.5, showpoints=true, linewidth=0.3mm](0.20000, 95.90047)(0.30000, 77.55919)(0.40000, 28.65116)(0.50000, 9.32818)(0.60000, 28.65116)
\endpsclip
\rput[t]{90}(0.13651, 7.50000){sum rate gain [\%]}
\rput[r](0.59365, 1.00000){SISO SNR 10dB, $\Np=2$}
\psframe[linecolor=black, fillstyle=solid, fillcolor=white, shadowcolor=lightgray, shadowsize=1mm, shadow=true](0.21270, 5.66038)(0.42222, 14.43396)
\rput[l](0.26349, 13.58491){DIS (sc.)}
\psline[linecolor=blue, linestyle=solid, linewidth=0.1mm](0.22540, 13.58491)(0.25079, 13.58491)
\rput[l](0.26349, 12.16981){DIS (spc.)}
\psline[linecolor=blue, linestyle=dashed, linewidth=0.3mm](0.22540, 12.16981)(0.25079, 12.16981)
\rput[l](0.26349, 10.75472){DIS (sc./spc.)}
\psline[linecolor=blue, linestyle=solid, linewidth=0.3mm](0.22540, 10.75472)(0.25079, 10.75472)
\psline[linecolor=blue, linestyle=solid, linewidth=0.3mm](0.22540, 10.75472)(0.25079, 10.75472)
\psdots[linecolor=blue, linestyle=solid, linewidth=0.3mm, dotstyle=square, dotscale=1.5 1.5, linecolor=blue](0.23810, 10.75472)
\rput[l](0.26349, 9.33962){DAS (sc.)}
\psline[linecolor=red, linestyle=solid, linewidth=0.1mm](0.22540, 9.33962)(0.25079, 9.33962)
\rput[l](0.26349, 7.92453){DAS (spc.)}
\psline[linecolor=red, linestyle=dashed, linewidth=0.3mm](0.22540, 7.92453)(0.25079, 7.92453)
\rput[l](0.26349, 6.50943){DAS (sc./spc.)}
\psline[linecolor=red, linestyle=solid, linewidth=0.3mm](0.22540, 6.50943)(0.25079, 6.50943)
\psline[linecolor=red, linestyle=solid, linewidth=0.3mm](0.22540, 6.50943)(0.25079, 6.50943)
\psdots[linecolor=red, linestyle=solid, linewidth=0.3mm, dotstyle=o, dotscale=1.5 1.5, linecolor=red](0.23810, 6.50943)

\end{pspicture}
\endgroup
 }
\caption{Gain through source coding (sc.) or superposition coding (spc.), maximized over all extents of backhaul.}
\label{f:ANALYSIS_SPCSWWZGAIN}
\end{figure}
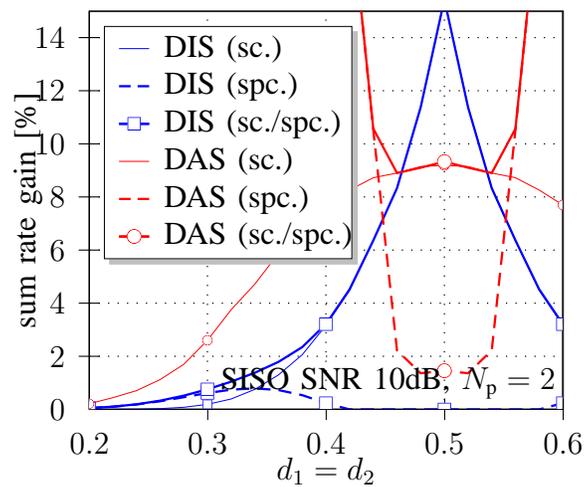

\begin{figure}
\centerline{\input{figure_ITANALYSIS_UL_M3K3_DISDAS_MONTECARLO1}}
\caption{Monte Carlo Results for $M=K=3$ and $\Nbs=2$ (cell-edge scenarios).}
\label{f:ANALYSIS_MONTECARLO1}
\end{figure}

\begin{figure}
\centerline{\input{figure_ITANALYSIS_UL_M3K3_DISDAS_MONTECARLO2}}
\caption{Monte Carlo Results for $M=K=3$ and $\Nbs=2$ (cell-center scenarios).}
\label{f:ANALYSIS_MONTECARLO2}
\end{figure}

%\begin{figure}
%\centerline{\input{figure_ITANALYSIS_UL_M3K3_ZEROINF_MCPC_DIAG_newsletter}}
%\caption{Monte Carlo Results for $M=K=3$ (cell-center scenarios).}
%\label{f:ANALYSIS_MONTECARLO2}
%\end{figure}
%
%\begin{figure}
%\centerline{\input{figure_ITANALYSIS_UL_M2K2_DISCIFDAS_LAMBDAAREA_AVGCHN_newsletter}}
%\caption{Best cooperation scheme as a function of UE locations, for a scenario with $M=K=2$, $\Nbs=2$ and $\beta=4$ bpcu.}
%\label{f:ANALYSIS_AREAPLOT}
%\end{figure}
%  
%\begin{figure}
%\centerline{\input{figure_ITANALYSIS_UL_M2K2_MCQDAS_RATEVSBH4_AVGCHN_newsletter}}
%\caption{Sum-rate as a function of backhaul for a channel of moderate, asymmetric interference.}
%\label{f:ANALYSIS_RATEVSBH2}
%\end{figure}
%  
%\begin{figure}
%\centerline{\input{figure_SYSTEMLEVEL_UL_IMPCSI_TPDISTRVSBH_newsletter}}
%\caption{Sum-rate as a function of backhaul for a channel of moderate, asymmetric interference.}
%\label{f:ANALYSIS_RATEVSBH2}
%\end{figure}
  
\begin{table*}[p]
	\centering
		\begin{tabular}{|l|c|c|c|c|} \hline
		\rowcolor[gray]{0.9} \cellcolor[gray]{0.8} & & & & \\
		\rowcolor[gray]{0.9} \cellcolor[gray]{0.8} \textbf{UPLINK} & \textbf{DIS} & \textbf{CIF} & \textbf{DAS-D} & \textbf{DAS-C} \\
		\rowcolor[gray]{0.9} \cellcolor[gray]{0.8} & & & & \\ \hline
		\cellcolor[gray]{0.9} & \multicolumn{3}{c|}{} & \\
		\cellcolor[gray]{0.9} \textbf{Decoding} & \multicolumn{3}{c|}{decentralized} & centralized \\
		\cellcolor[gray]{0.9} & \multicolumn{3}{c|}{} & \\ \hline
		\cellcolor[gray]{0.9} & & & \multicolumn{2}{c|}{} \\
		\cellcolor[gray]{0.9} \textbf{Exchanged} & decoded & quantized & \multicolumn{2}{c|}{quantized} \\
		\cellcolor[gray]{0.9} \textbf{signals} & messages & sequences & \multicolumn{2}{c|}{receive signals} \\
		\cellcolor[gray]{0.9} & & & \multicolumn{2}{c|}{} \\ \hline
		\cellcolor[gray]{0.9} & & & & \\
		\cellcolor[gray]{0.9} \textbf{Achievable} & SIC gain & partial & array + & array + \\
		\cellcolor[gray]{0.9} \textbf{gains} & (interference & interference & spat. mult. & spat. mult. + \\
		\cellcolor[gray]{0.9} & cancellation) & cancellation & gain & SIC gain \\
		\cellcolor[gray]{0.9} & & & & \\ \hline
		\cellcolor[gray]{0.9} & weak, asymm. & very weak, & very weak, & strong \\
		\cellcolor[gray]{0.9} \textbf{Suitable in} & interference & asymm. interf. & symm. interf. & interference \\
		\cellcolor[gray]{0.9} \textbf{scenarios} & low backh. & low backh. & low backh. & large backh. \\
		\cellcolor[gray]{0.9} & high SNR & high SNR & low SNR & low SNR \\ \hline
		\cellcolor[gray]{0.9} & & \multicolumn{2}{c|}{} & \\
		\cellcolor[gray]{0.9} \textbf{Source} & provide little & \multicolumn{2}{c|}{provide little gain, } & major potential \\
		\cellcolor[gray]{0.9} \textbf{coding} & gain, but possible & \multicolumn{2}{c|}{and are highly questionable} & gains, \\
		\cellcolor[gray]{0.9} \textbf{concepts} & if interference & \multicolumn{2}{c|}{from implementation} & but highly \\
		\cellcolor[gray]{0.9} & is also decoded & \multicolumn{2}{c|}{point of view} & questionable \\
		\cellcolor[gray]{0.9} & & \multicolumn{2}{c|}{} & \\ \hline
		\cellcolor[gray]{0.9} & \multicolumn{2}{c|}{} & & \\
		\cellcolor[gray]{0.9} \textbf{Channel} & \multicolumn{2}{c|}{local knowledge from} & global CSI & global CSI \\
		\cellcolor[gray]{0.9} \textbf{knowl. req.} & \multicolumn{2}{c|}{each BS to all UEs sufficient} & at {\em all} BSs & at {\em one} BS \\
		\cellcolor[gray]{0.9} & \multicolumn{2}{c|}{} & & \\ \hline
		\cellcolor[gray]{0.9} & {\em moderate}, if & & & \\
		\cellcolor[gray]{0.9} & interf. only needs & & & {\em high}, as \\
		\cellcolor[gray]{0.9} & re-encoding / SIC & {\em low}, due to & {\em low}, as only & all UEs are \\
		\cellcolor[gray]{0.9} \textbf{Complexity} & (w/o src. coding),& simple signal & one UE is & successively \\
		\cellcolor[gray]{0.9} & {\em high} if dec. of & subtraction & decoded & or jointly \\
		\cellcolor[gray]{0.9} & mult. UEs / SIC & & & decoded + SIC \\
		\cellcolor[gray]{0.9} & (w/ src. coding) & & & \\ \hline
		\end{tabular}
		\caption{Comparison of uplink BS cooperation schemes, considering practical aspects.}
			\label{t:PRACTICAL_SUMMARY}	
\end{table*}

  \clearpage
\end{document}